\documentclass[fleqn,usenatbib]{mnras}

\usepackage{graphicx}	
\usepackage{amsmath}	
\usepackage{amssymb}	

\newcommand{\pulsarone}{J1810$+$1744}
\newcommand{\pulsartwo}{J1744$-$1134}
\newcommand{\pulsarthree}{J1231$-$1411}
\newcommand{\pulsarfour}{J2256$-$1024}
\newcommand{\pulsarfive}{J1723$-$2837}
\newcommand{\pulsarsix}{J1400$-$1431}
\newcommand{\pulsareight}{J0952$-$0607}
\newcommand{\pulsarnine}{J2241$-$5236}
\newcommand{\pulsarten}{J0437$-$4715}

\title{Strengthening the bounds on the r-mode amplitude with X-ray observations of millisecond pulsars}

\author[T. Boztepe et al.]{
Tu\u{g}ba Boztepe,$^{1}$\thanks{E-mail: tugbabztp@gmail.com}
Ersin G\"o\u{g}\"u\c{s}$^{2}$,
Tolga, G\"uver$^{3,4}$,
Kai Schwenzer$^{3}$\\
$^{1}$Istanbul University, Graduate School of Sciences, Department of Astronomy and Space Sciences, 34116, Beyaz\i t, Istanbul, Turkey\\ 
$^{2}$Faculty of Engineering and Natural Sciences, Sabanc\i~University, Orhanl\i-Tuzla 34956, Istanbul, Turkey \\
$^{3}$Istanbul University, Science Faculty, Department of Astronomy and Space Sciences, Beyaz\i t, 34119, Istanbul, Turkey\\
$^{4}$Istanbul University Observatory Research and Application Center, Istanbul University 34119, Istanbul Turkey}

\begin{document}
\label{firstpage}
\maketitle

\begin{abstract}
r$-$mode oscillations have been shown to have a significant potential to constrain the composition of fast spinning neutron stars. Due to their high rotation rates, millisecond  pulsars (MSPs) provide a unique platform to constrain the properties of such oscillations, if their surface temperatures can be inferred. 
We present the results of our investigations of archival X-ray data of a number of MSPs, as well as recent {\em XMM-Newton} observations of PSR~\pulsarone ~and PSR~\pulsarnine. Using the neutron star atmosphere (NSA) model and taking into account various uncertainties, we present new bounds on the surface temperature of these sources. Thereby we significantly strengthen previous bounds on the amplitude of the r-mode oscillations in millisecond pulsars and find rigorous values as low as $\alpha\lesssim 3 \times 10^{-9}$. 
This is by now about three orders of magnitude below what standard saturation mechanisms in neutron stars could provide, which requires very strong dissipation in the interior, strongly pointing towards a structurally complex or exotic composition of these sources.
At such low temperatures sources could even be outside of the instability region, and taking into account the various uncertainties we obtain for an observed surface temperature a simple frequency bound below which r-modes are excluded in slower spinning pulsars. 

\end{abstract}




\section{Introduction}

r$-$mode asteroseismology provides a unique opportunity to probe the opaque compact star interior. r$-$modes \citep{Papaloizou1978,Andersson1998,Lindblom1998,Friedman1998, Andersson2000} are quasi-toroidal oscillations in rotating stars that occur owing to the Coriolis effect. Because r-modes are unstable \citep{Andersson1998} due to the Chandrasekhar-Friedman-Schutz (CFS) mechanism \citep{Chandrasekhar1970, Friedman1978}, they would spontaneously arise under the emission of gravitational waves and spin the star down, unless the instability can be tamed by dissipation within the star, which depends sensitively on the composition. 
r-modes are damped in slowly spinning sources, but should be unstable in ordinary neutron stars spinning with millisecond frequencies over a range of typical temperatures, since in this instability region viscous dissipation cannot prevent their instability to gravitational wave emission. In this case the amplitude would be saturated by non-linear, amplitude-dependent enhancement and would strongly heat the star if the amplitude becomes large.
Without enhanced damping mechanisms millisecond pulsars would be expected to be trapped inside the instability region \cite{Alford2014} due to the substantial heating of even small-amplitude r-modes as well as their tiny observed spindown rates.
Therefore, bounds on the temperatures of these sources conversely set bounds on the amplitude of potential r-modes in observed sources.
This has previously been shown for accreting sources in Low Mass X-ray Binaries (LMXBs) \citep{Mahmoodifar2013,Alford2014}, as well as millisecond pulsars (MSPs) \citep{Alford2013,Schwenzer,Chugunov:2017,Bhattacharya2017,Mahmoodifar2017,Ho:2019myl}, and led to very low bounds on the r-mode amplitude of $\alpha\lesssim10^{-8}$.

The dissipation and cooling, that determines the steady state temperature, depends strongly on the matter inside the star. Measurements or bounds on surface temperatures of non-accreting MSPs in turn impose bounds on the r-mode amplitude, since these sources would be hotter if the r-mode amplitude would be higher. However, for non-accreting MSPs detailed X-ray measurements are still limited due to their lower flux. 
An actual surface temperature measurement is only available for the closest sources such as PSR~\pulsarten ~ \cite{Zavlin2002,Durant:2011je,Gonzalez:2019} and PSR J2124$-$3358 \cite{Rangelov:2016syg}.
Besides the observed surface temperature, the spin frequency has been shown to be the decisive factor for the amplitude bounds \citep{Mahmoodifar2013,Schwenzer}.

In this paper, we report new results on thermal bounds of fast-spinning MSPs. We also present results of two new {\em XMM-Newton} observations of the $602\,{\rm Hz}$ pulsar PSR~\pulsarone~and the $457.31\,{\rm Hz}$ pulsar PSR~\pulsarnine~allowing us to perform the first detailed spectral analysis of these sources. Moreover, we report the re-analysis of archival data of several other MSPs, including an update of our earlier analysis for PSR \pulsarthree ~\citep{Schwenzer} and the fastest spinning pulsar in the Galactic field PSR~\pulsareight. We use here an improved version of the method we introduced in \citep{Schwenzer} to take into account  dominant microscopic and macroscopic (astrophysical) uncertainties in the analysis, e.g., the distance, to obtain stringent upper limits. Moreover, we now employ the more realistic neutron star atmosphere (NSA) model, appropriate for  old systems, instead of assuming a mere {\em blackbody}. Based on the new spectral analyses we improve the previous r-mode amplitude bounds for these sources and even within uncertainties we now have several sources with strict upper limits of a few times $10^{-9}$. 

The structure of the paper is as follows. After providing some introductory information on the individual sources we use in this study in Section~\ref{sec:source}, we provide the details of the observations and data analysis~in Section~\ref{sec:analysis}. We present the limits on thermal emission obtained from the X-ray data of the sources in Section~\ref{sec:limits}.
We sketch the thermal impact of r$-$modes and derive the resulting r-mode amplitude bounds in Section~\ref{sec:rmode}. In Section~\ref{sec:prermode} we derive a general condition if r$-$modes can be present in observed millisecond sources. 
Finally, we conclude and discuss our results from X-ray observations in Section~\ref{sec:discus}.

\section{Millisecond Pulsars Investigated}\label{sec:source}

We provide below a brief introduction to each neutron star system investigated in our study.

\subsection{\em PSR~\pulsarone:}
\pulsarone ~was discovered in a 350~MHz targeted Green Bank Telescope (GBT) search of unidentified Fermi point sources \cite{Hessels2011}, and has been identified as a canonical Black-Widow \cite{Breton2013}. 
The dispersion measure (DM) is measured to be 39.7~pc~$\rm cm^{-3}$ \cite{Breton2013}, which corresponds to a distance of 2~kpc (using the NE2001 model). The pulsar has a spin period $P_{\rm spin}$ = 1.66~ms. 
The pulsar has a very small mass companion M$_{\rm min}^{c}$ = 0.045~$M_\odot$ orbiting with a period of $P_{\rm orb}$ = 3.6~hr. At low radio frequencies, this source is one of the brightest known millisecond pulsars \cite{Polzin2018}. 

Previous to our new {\em XMM-Newton} observation presented in detail below, \cite{Gentile2014} fit the only available X-ray spectrum, with a combination of a thermal and a non-thermal component with fixed temperature and photon index values (kT = $150$~eV, $\Gamma=1.5$). The combined flux is measured as $F_{\rm total} = 2.0_{-0.6}^{+0.5}\times10^{-14}~{\rm{erg}}\,{\rm {cm}^{-2}\,{\rm{s}^{-1}}}$, where the flux of the individual components are calculated as, 
$F_{\rm bb} = 0.7_{-0.4}^{+0.4}\times10^{-14}~{\rm{erg}}\,{\rm {cm}^{-2}\,{\rm{s}^{-1}}}$, 
$F_{\rm pow} = 2.0_{-0.7}^{+0.6}\times10^{-14}~{\rm{erg}}\,{\rm {cm}^{-2}\,{\rm{s}^{-1}}}$.

\subsection{\em PSR~\pulsartwo:}
\pulsartwo ~was discovered as part of the Parkes 436~MHz survey of the southern sky \cite{Bailes1997}. The pulsar has a spin period $P_{\rm spin}$ = 4.075~ms and a period derivative of $\dot P_{\rm int}$ = $0.86\times10^{-20} \rm s~s^{-1}$ \cite{Manchester2013}. 
Recently, using the pulsar timing model, \cite{Perera2019}~reported a precise parallax measurement for \pulsartwo, $\pi(mas)$ = 2.44, this resulting in a very precise distance of $d$ = $0.410\pm{0.008}~{\rm kpc}$.


\subsection{\em PSR~\pulsarthree:} 
\pulsarthree ~was discovered by Fermi~LAT as one of the brightest gamma-ray MSPs in the sky ($F_{100}$ = $10.57_{-0.32}^{+0.62}\times10^{-8}~\rm erg~cm^{-2}~s^{-1}$), it has a spin period of $P$ = 3.684~ms and a period derivative $\dot P_{\rm }$ = $2.26\times~10^{-20}~\rm{s}~\rm{s}^{-1}$. 
This source is in a binary system which has an orbital period of 1.8601~d, with a dispersion measure (DM) of 8.09~pc~$\rm cm^{-3}$ \cite{Bassa2016}. 
The observed period derivative implies a surface magnetic field strength of ($2-3)\times10^{8}$~G and a spin-down luminosity of $\sim2\times10^{34}\,{\rm erg \, s}^{-1}$ \cite{Ransom2011}. 

The flux of the source in the $0.5-3.0$~keV range was measured as ($1.15 \pm 0.05) \times 10^{-13} \rm erg \, cm^{-2} s^{-1}$.
In addition, we previously modeled the {\em XMM-Newton} data with a {\em blackbody}, giving an unabsorbed flux of $F_{\rm bb}$ = $15_{-7.4}^{+5.3}\times 10^{-14}~\rm erg~cm^{-2}~s^{-1}$ in the 0.5-8~keV band, and when taking into account a distance of 0.4~kpc an X-ray luminosity of $L_{\rm bb}$ = $2.9_{-1.4}^{+1.0}\times10^{30}~\rm erg~s^{-1}$. Alternatively, the observed spectrum can be modeled with a {\em Power-law} with a slope $\Gamma$ = $4.23_{-0.38}^{+0.41}$, and the column density is calculated as $N_{\rm H} = 1.8_{-0.5}^{+0.6}\times 10^{21}{\rm cm}^{-2}$ \cite{Schwenzer}. 

\subsection{\em PSR~\pulsarfour:} 
\pulsarfour ~was discovered during the 350~Mhz GBT pulsar drift scan survey \cite{Boyles2011}. This pulsar is a Black-Widow with a degenerate companion of mass $0.1~M_{\odot}$, which exhibits radio eclipses. It has a spin period of 2.29~ms and a short orbital period of 5.1~hr \cite{Gentile2018}. This source has a DM of 13.8~${\rm pc}~{\rm cm}^{-3}$, indicating a distance of $\approx$ 0.6~kpc, using the NE2001 model \cite{CordesLazio2002} or using the YMW16 model the same dispersion measure indicates a distance of 1.33~kpc.

The combined flux was measured as $F_{X}$ = $4.6_{-1.6}^{+2.5}\times10^{-14}~\rm erg~cm^{-2}~s^{-1}$, where the flux of the single components were calculated as, $F_{\rm pow}$ = ($5.3\pm0.6)\times10^{-14}~\rm erg~cm^{-2}~s^{-1}$, $F_{\rm bb}$ = $3.2_{-1.6}^{+2.6}\times10^{-14}~\rm erg~cm^{-2}~s^{-1}$ by \cite{Gentile2014}.

\subsection{\em PSR~\pulsarfive:} 
\pulsarfive ~was discovered by \cite{Faulkner2004} during the Parkes Multi-beam survey, it has a spin period of 1.86~ms and the measured spin-down rate is $\sim7.5 \times 10^{-21}~\rm s~s^{-1}$. This pulsar is a Redback millisecond pulsar with a low-mass companion in a 14.8~hr orbit. The pulsar's DM is 19.69~pc ${\rm cm^{-3}}$ indicating a distance of 0.75~kpc \cite{Crawford2013}, using the NE2001 model \cite{CordesLazio2002}. However, GAIA's second data release \cite{Jennings2018} yielded a very precise distance measurement for this source. Based on GAIA results, \cite{Jennings2018} calculated the distance to be $d$ = $0.91_{-0.04}^{+0.05}$~kpc. 
In this paper we will use this new model independent distance measurement for further analysis. 

The X-ray spectrum of this source could be modeled with an absorbed {\em Power-law} model by \cite{Kong2017}, based on NuSTAR observations. The best-fit model parameters inferred are $N_{\rm H} = 3.9_{-2.0}^{+2.7}\times 10^{21}{\rm cm}^{-2}$ and $\Gamma$ = $1.28 \pm 0.04$ and as a consequence the X-ray flux was calculated as $F_{X}$=$(9.6{\pm0.5})\times10^{-12}~\rm erg~s^{-1}~cm^{-2}$ \cite{Kong2017}.

\subsection{\em PSR~\pulsarsix:} 
\pulsarsix ~was discovered by the Pulsar Search Collaboratory in the Green Bank 350 MHz Drift Scan Survey \cite{Rosen2013}. This pulsar has a dispersion measure of 4.9~pc~${\rm cm}^{-3}$. Its spin period is 3.08~ms with a period derivative $\dot P_{\rm }$ = $7.23\times~10^{-21}~\rm{s}~\rm{s}^{-1}$. The binary orbital period is found to be 9.5~d, with a white dwarf companion mass $M_{c,{\rm min}}$ $\sim0.26$ $M_\odot$ \citep{Swiggum2017}. The X-ray luminosity of the source is reported as $L$ = $10^{29} \rm erg ~s^{-1}$ in the $0.3-10.0$~keV range for a distance of 270~pc, and the spindown luminosity is $<\,3\times10^{33}~{\rm erg}~{\rm s}^{-1}$. 

The inferred spectral parameters of a {\em blackbody} model are $kT$ = ($0.15 \pm0.2)\, {\rm keV}$, with an apparent radius of $R = 0.06_{-0.04}^{+0.05}~{\rm km}$, an unabsorbed flux of $F_{abs}$ = ($1.07 \pm 0.15) \times 10^{-14}~{\rm erg}~{\rm cm}^{-2}~{\rm s}^{-1}$, for a fixed value of the atomic hydrogen column density, $N_{\rm H} = 1.5\times10^{20} \rm{cm}^{-2}$ \citep{Swiggum2017}.  

\subsection{\em PSR~\pulsareight:} 
\pulsareight ~was discovered by the Low-Frequency Array (LOFAR) survey at 135 MHz. It is in a 6.42~hr binary with a very low-mass companion and it is the fastest-spinning known pulsar in the galactic field ($707\,{\rm Hz}$) \cite{Bassa2017}. The observed properties of the system show that it is a Black-Widow binary. The DM is reported to be 22.412~pc ${\rm cm}~{\rm s}^{-3}$
and a period derivative $\dot P_{\rm }$ = $4.6\times~10^{-21}~\rm{s}~\rm{s}^{-1}$ \cite{Ho:2019myl}. This value is used to infer two different distances assuming the YMW16 and the NE2001 models as $d=1.74\,{\rm kpc}$ and $ d= 0.97\,{\rm kpc}$ \cite{Bassa2017}, respectively. 

\cite{Ho:2019myl} recently presented an analysis of the {\em XMM-Newton} observation of the system.  The spectral fit, using a {\em Power-law}, results in a photon index of $\Gamma$ $\approx 2.55_{-0.4}^{+0.5}$ and $f_{unabs}$ = $9\times10^{-15}~\rm erg~s^{-1}~cm^{-2}$ in the 0.3$-$10~keV range.
Their result for the luminosity $L_{X}$ = $3\times10^{30} \rm erg \, s^{-1}$, assuming a distance of 1.74~kpc, is ten times lower than the upper limits set using a short Swift exposure by \cite{Bassa2017}. Note that the flux result obtained using a single {\em Power-law} is consistent with the value we find here (see Table \ref{t_sp_res}).

\subsection{\em PSR~\pulsarnine:}
\pulsarnine~ was discovered by Fermi$-$LAT \cite{Keth2011}. It has a spin period 2.19~ms and a period derivative $\dot P_{\rm }$ = 6.6$\times~10^{-21}~\rm{s}~\rm{s}^{-1}$. Using these timing measurements of the pulsar the characteristic age, the surface magnetic field strength and the rate of energy loss has been derived as $\tau = 5\times~10^{9}$~yr, B~=~1.2$\times~10^{8}$~G, and $\dot{E}=2.5\times~10^{34}~\rm{erg} ~\rm{s}^{-1}$, respectively. The pulsar has a very low mass companion with an orbital period of 3.5 hours. 

This source has a low dispersion measure of 11.41~pc~${\rm cm}^{-3}$, indicating a distance of $\approx$ 0.49~pc (using the NE2001 model \cite{CordesLazio2002}) \citep{Keth2011}, whereas assuming the YMW16 model \citep{Yao2017}, this dispersion measure results in a distance of 0.96~kpc. Previous to our new {\em XMM-Newton} observation discussed below, the X-ray spectrum of this source was fitted with an absorbed single {\em blackbody} model with a temperature, kT =0.26$\pm$0.04~keV \cite{Keth2011}. 
Using the Galactic hydrogen column density, $\rm{N}_{\rm{H}}= 1.21\times~10^{20}~\rm{cm}^{-2}$, \cite{Kalberla2005} and assuming a distance of 0.49~kpc, the X-ray luminosity is measured as $\sim2\times~10^{30}~{\rm erg} {\rm s}^{-1}$, which is less  than 1\% of the rotational energy loss rate \cite{Keth2011}.

\section{Observations and Data Analysis}\label{sec:analysis}

We concentrate in this work on the X-ray observations of eight millisecond pulsars which are rapidly rotating, nearby and high quality X-ray observations could be found. These pulsars are introduced in Section \ref{sec:source} and a log of their observations used here, is given in Table \ref{t_obs}. For two pulsars (PSR~\pulsarone~ and PSR~\pulsarnine), we obtained new {\em XMM-Newton} observations and provide here the first spectral results of these data. {\em XMM-Newton} observed PSR~\pulsarone~ for 82~ks on 16th October 2017 and PSR~\pulsarnine~ for 52~ks on 28th November 2018, with ObsIDs, 0800880201 and 0824230201 respectively (see Table \ref{t_obs}).

Both the Chandra X-ray Observatory~(CXO) and {\em XMM-Newton} observatories' data presented in this paper were analyzed using the standard scientific analysis software for each satellite. For the CXO data analysis, we used CIAO\footnote{\url{http://cxc.cfa.harvard.edu/ciao/}} version 4.9 with CALDB version 4.7.3. We used the {\em chandra\_repro} tool to create calibrated Level~2 event files and the {\em spec\_extract} tool to extract source and background X-ray spectra and generate appropriate response and ancillary response files. We used circular regions with typical radii of 2\arcsec-3\arcsec and 8\arcsec-15\arcsec~radius, for source and background, respectively. 

In a similar way, all the data analysis and calibration regarding the XMM-{\it Newton} data has been performed using SAS\footnote{\url{https://www.cosmos.esa.int/web/xmm-newton}} version 20160201\_1833-15.0.0 with the most up to date calibration files as of 1st of June 2018. We used the {\em epproc} and {\em emproc} tools to create calibrated event files for the European Photon Imaging Camera (EPIC) pn, MOS1 and MOS2. In general we had to use pn camera data, which resulted in a higher signal. However whenever possible we also used MOS data as well. Using the {\em espfilt} tool, we investigated carefully the existence of soft proton contamination. In some cases this resulted in a decrease in the exposure times of the X-ray spectra; the total exposure times are given in Table \ref{t_obs}. For example, in the case of PSR~\pulsarone~the most recent observation was obtained by XMM-{\it Newton} on October 16, 2017~(ObsID: 0800880201) for a total exposure time of 82~ks, however, unfortunately it was heavily affected by solar particle background and therefore it was only possible to extract 24 and 32~ks of data from the EPIC~pn and MOS, respectively. For all the XMM-{\it Newton} data, the source and appropriate background spectra were extracted from circular regions of roughly 640 and 1200 pixel radius from the same CCD chip. 

\begin{figure*}
\includegraphics[scale=0.3,angle=270]{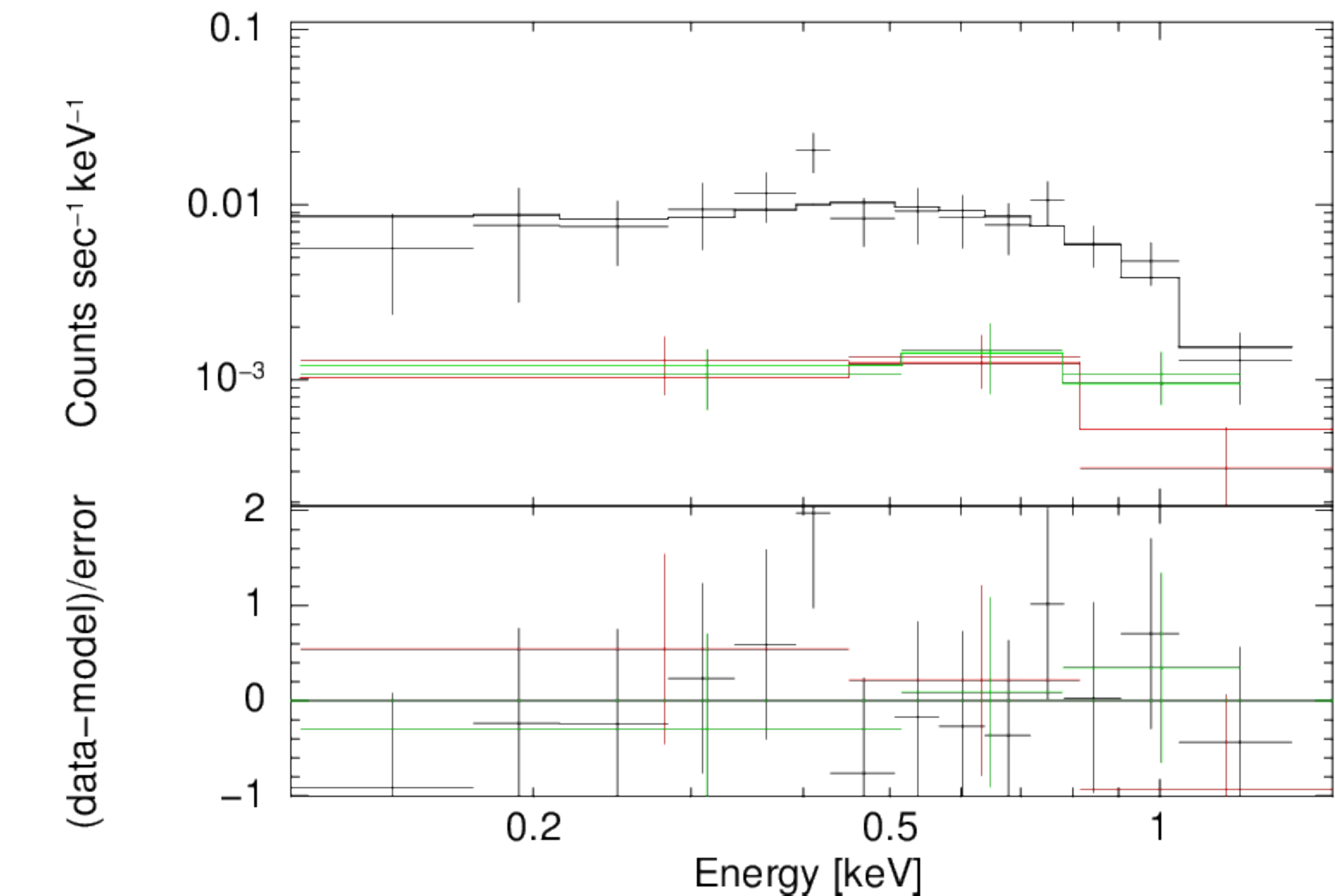}
\includegraphics[scale=0.3,angle=270]{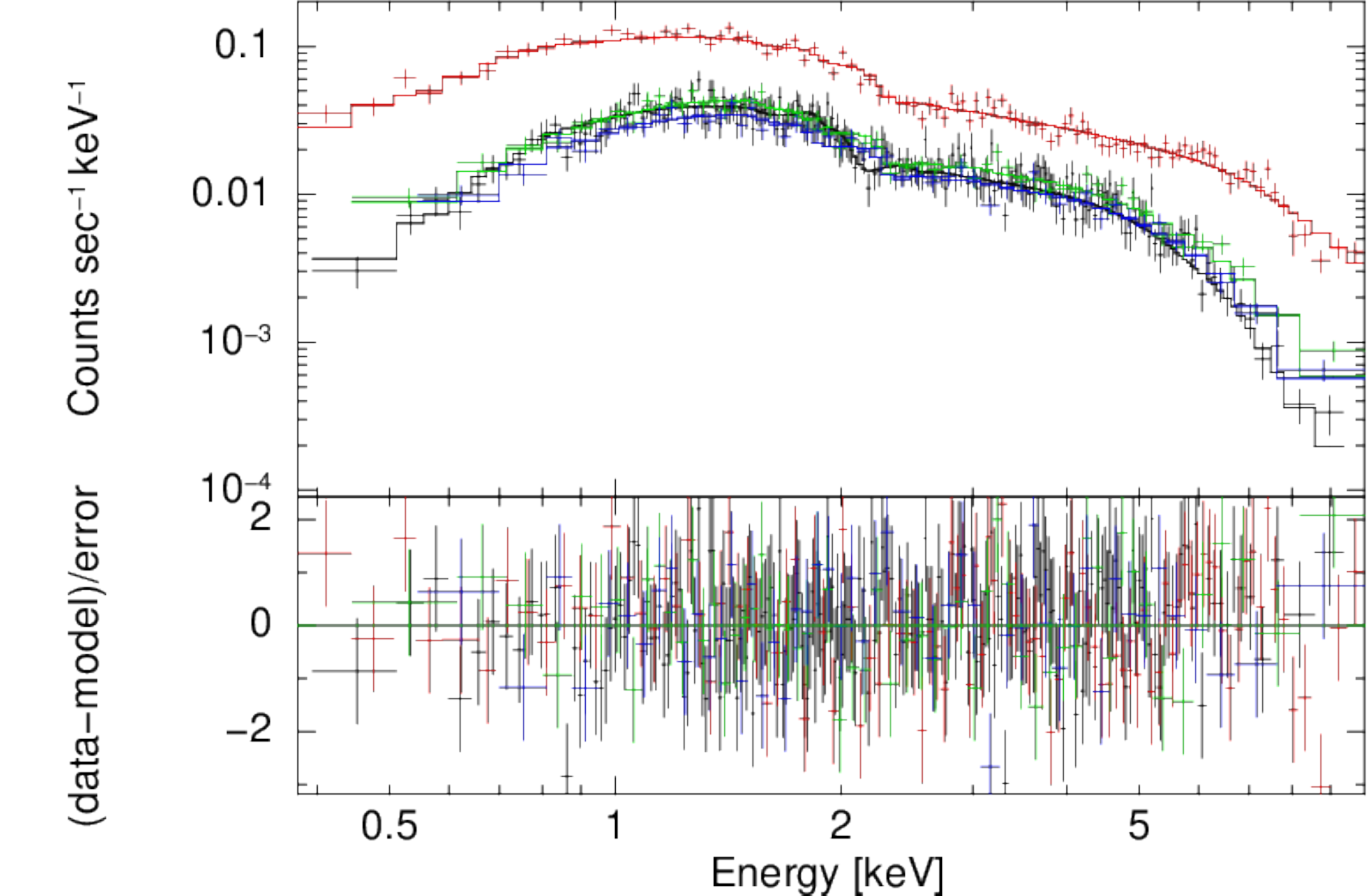}
\caption{\label{fig:sp1} X-ray spectra of PSR \pulsarsix~(left panel)~and PSR \pulsarfive~(right panel) together with the best fit model. Lower panels show the residuals from the model in units of the statistical uncertainty of the data. Colors of the labels in the upper panels denote the corresponding data in Chandra, {\em XMM-Newton}, Swift and {\em Suzaku} data are shown by different colors.}
 
\end{figure*}

\begin{figure*}
\includegraphics[scale=0.3,angle=270]{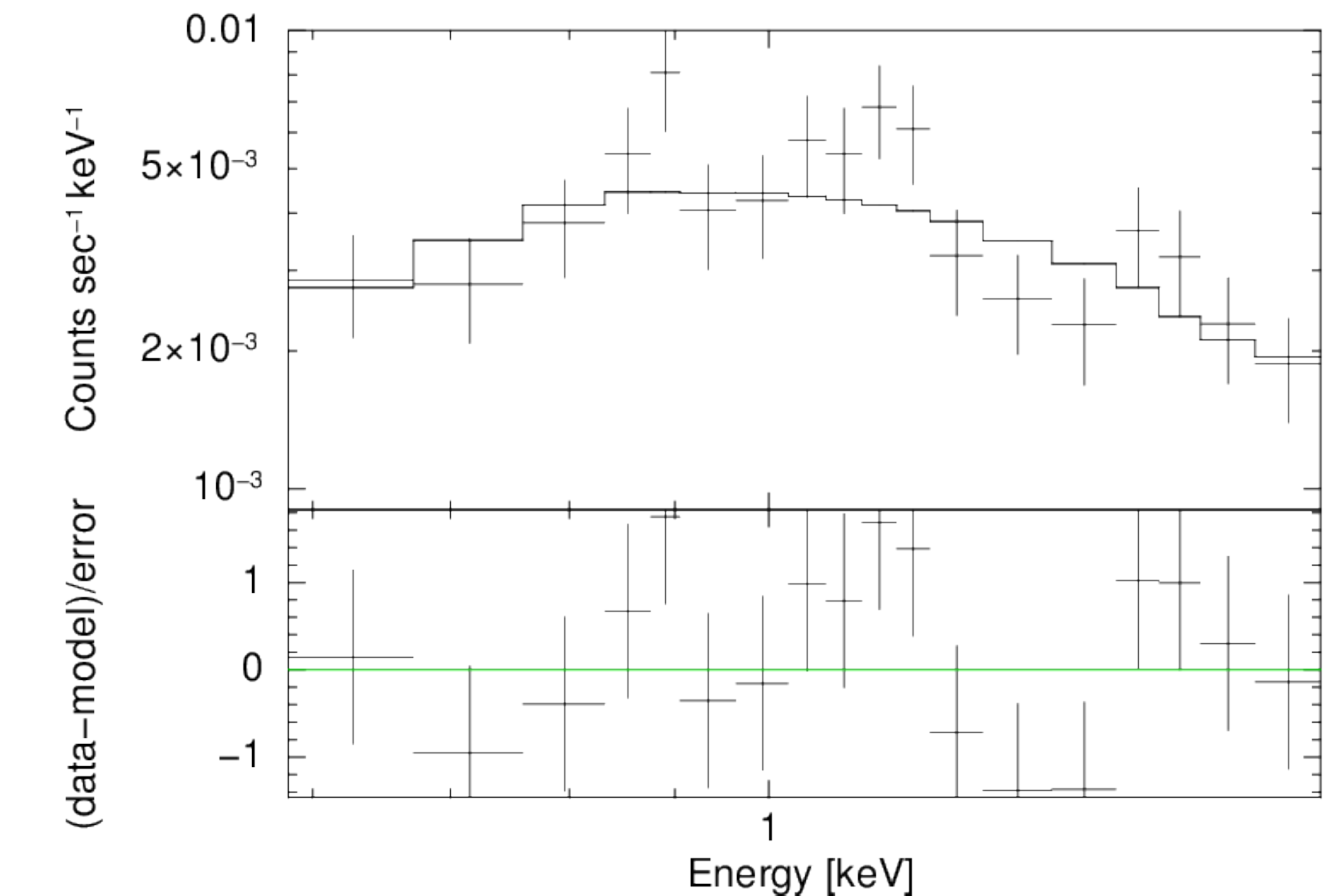}
\includegraphics[scale=0.35, angle=270]{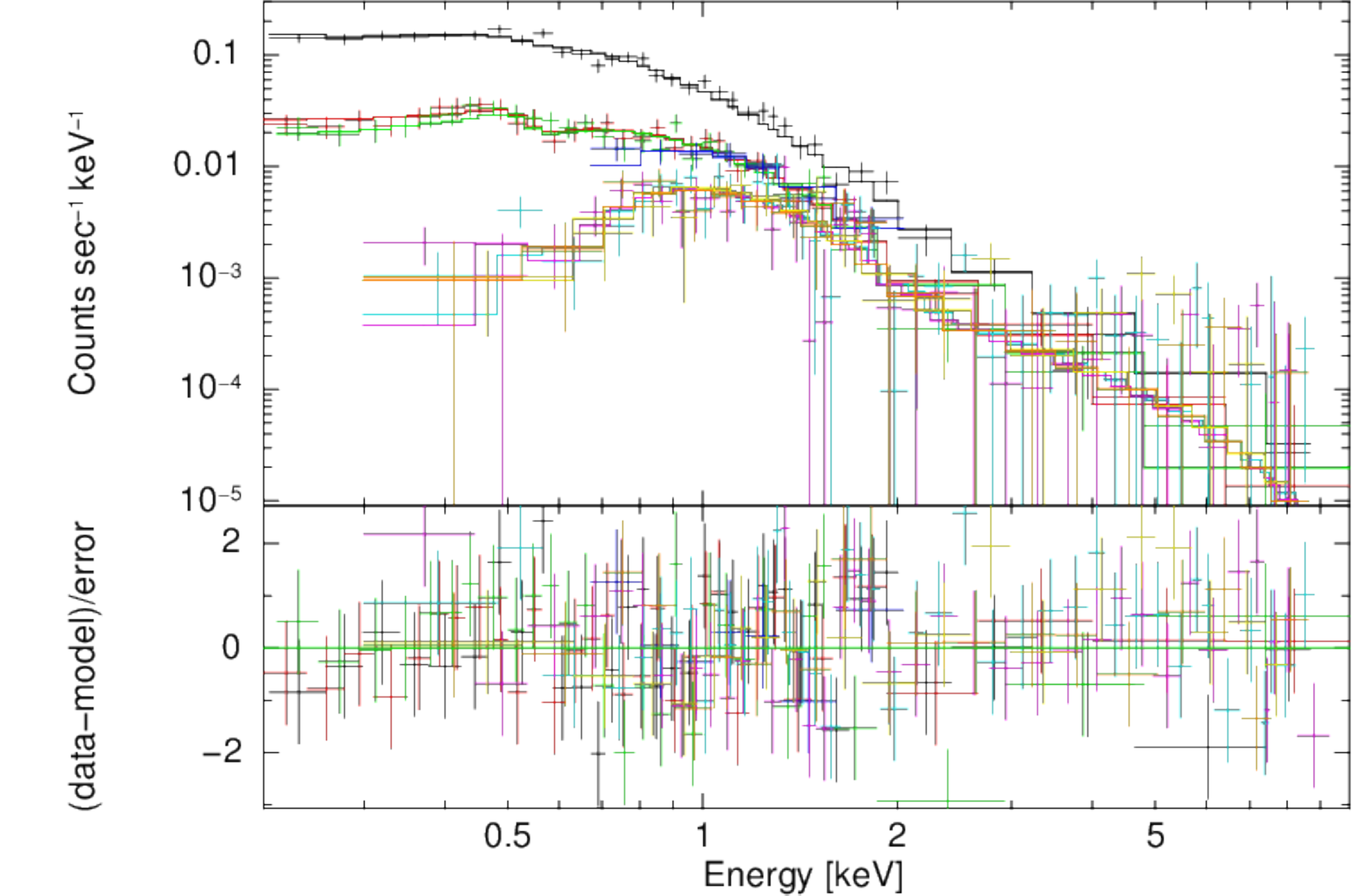}
\caption{\label{fig:sp2} As in Figure \ref{fig:sp1} for PSR \pulsartwo~(left panel)~and PSR \pulsarthree~(right panel). Colors of the labels in the upper panels denote the corresponding data in Chandra, {\em XMM-Newton}, Swift and {\em Suzaku}. The Chandra, {\em XMM-Newton}, Swift and {\em Suzaku} data are shown by different colors.
}
\end{figure*}

\begin{figure*}
\includegraphics[scale=0.3, angle=270]{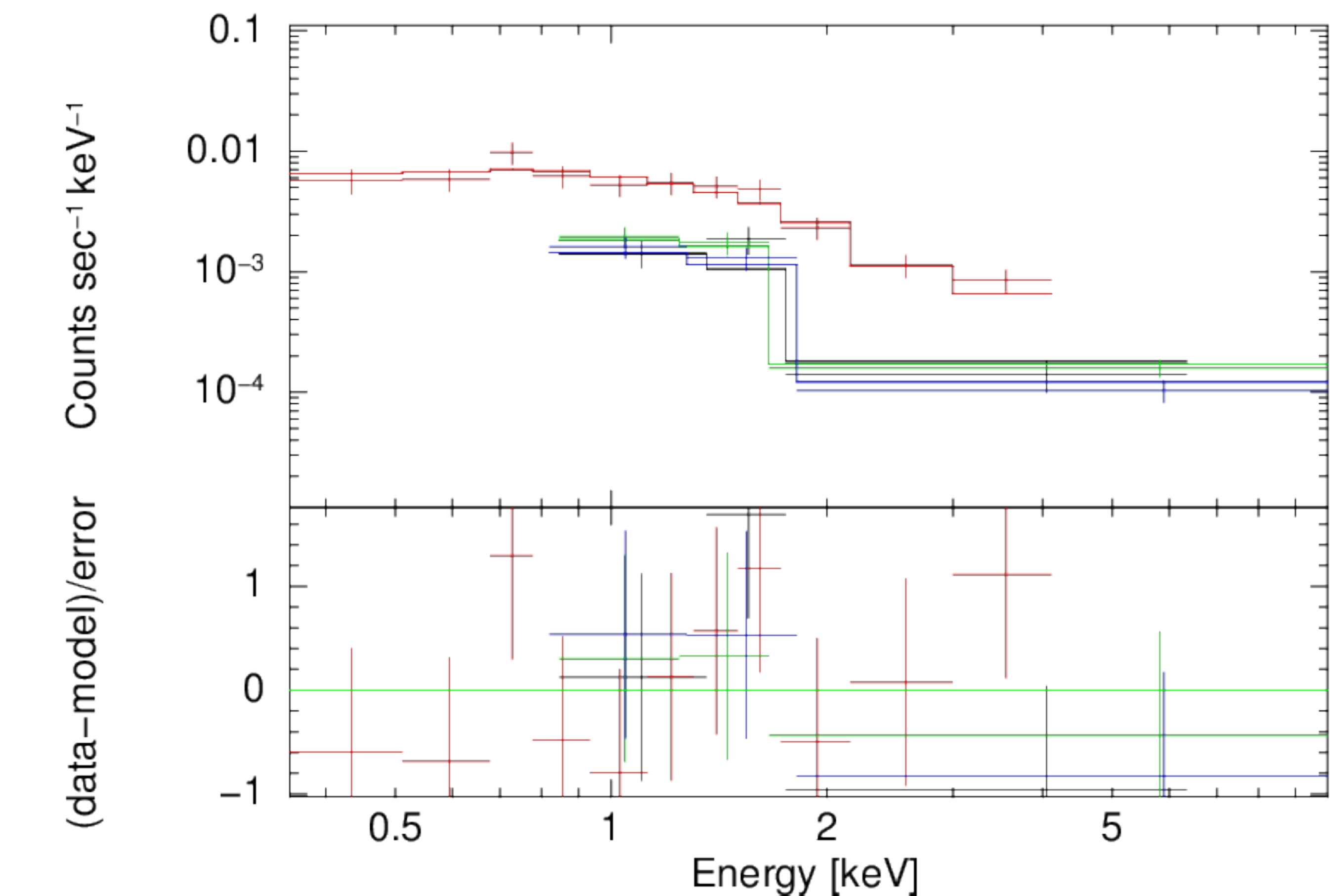}
\includegraphics[scale=0.3, angle=270]{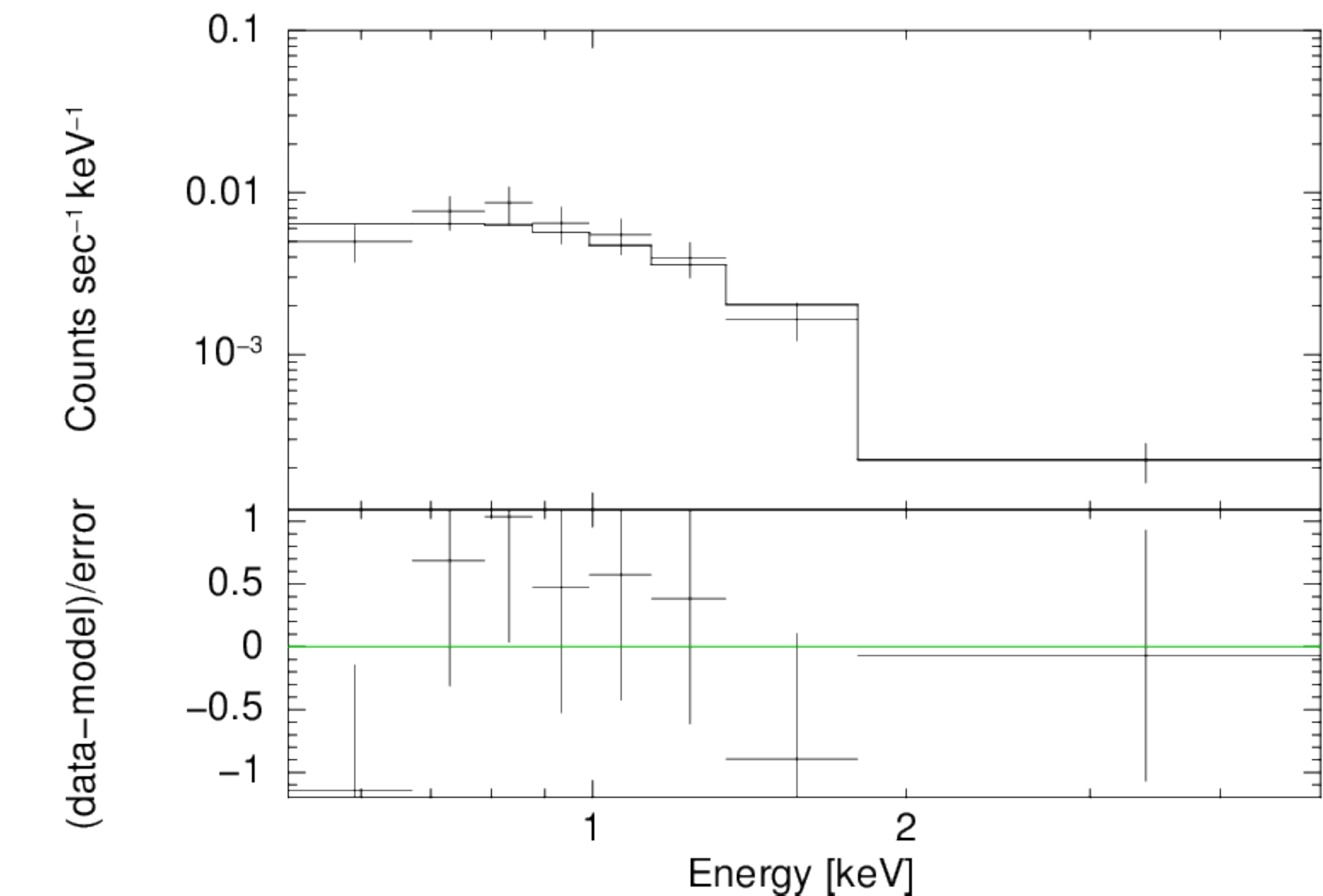}
\caption{\label{fig:sp3} As in Figure \ref{fig:sp1} for PSR \pulsarone~(left panel)~and PSR \pulsarfour~(right panel). Colors of the labels in the upper panels denote the corresponding data in Chandra, {\em XMM-Newton}, Swift and {\em Suzaku}. The Chandra, {\em XMM-Newton}, Swift and {\em Suzaku} data are shown by different colors.}
\end{figure*}

\begin{figure*}
\includegraphics[scale=0.3, angle=270]{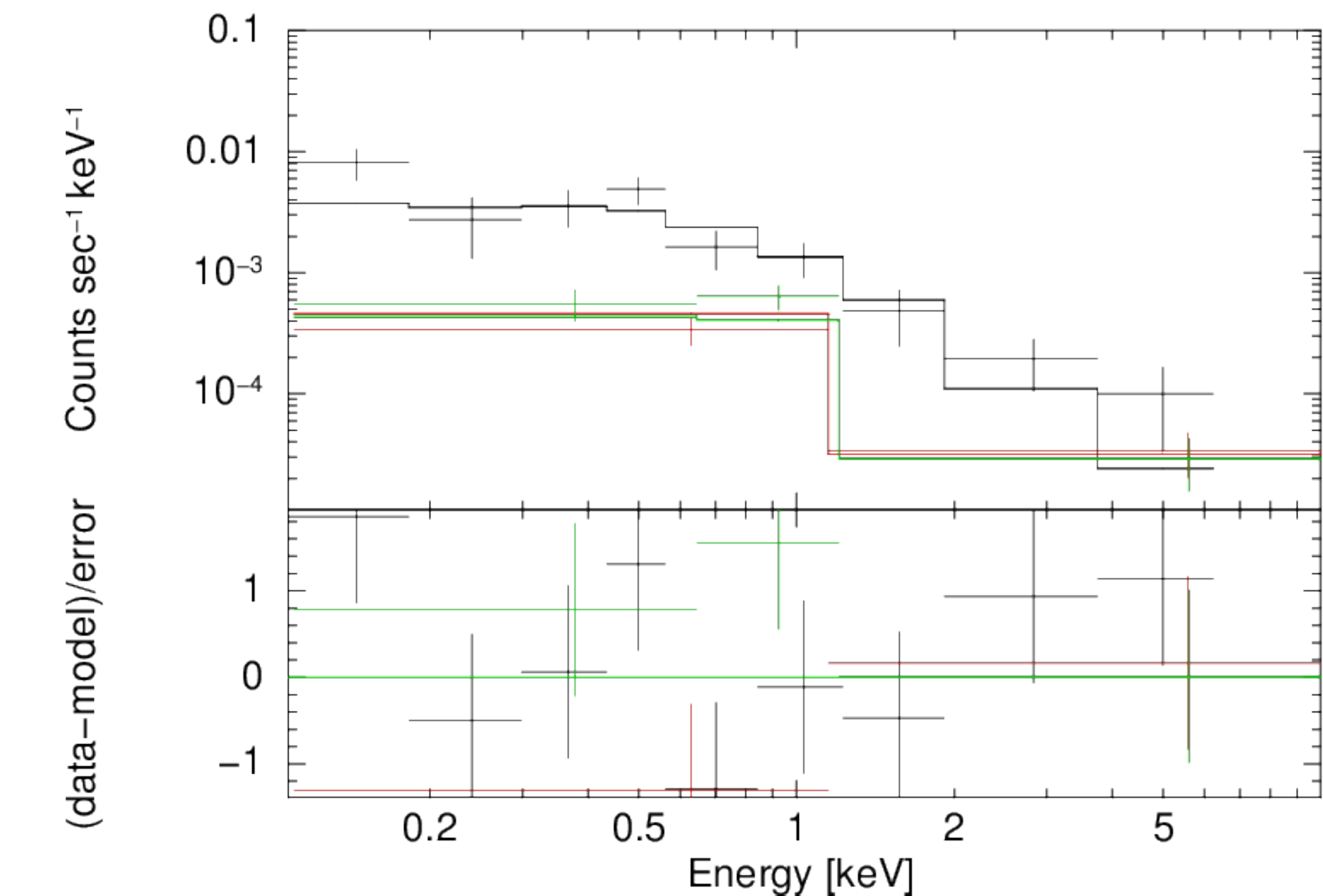}
\includegraphics[scale=0.3, angle=270]{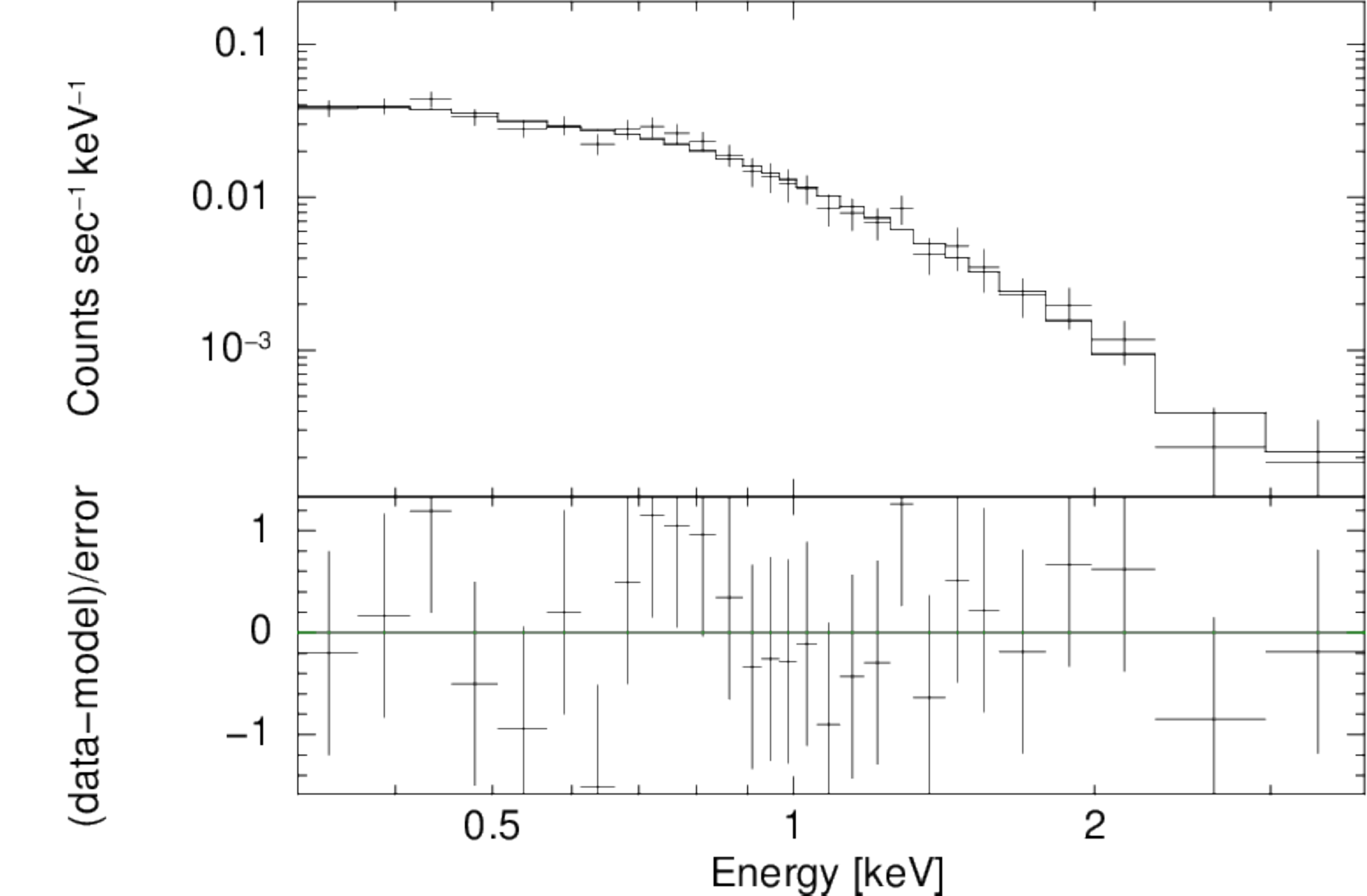}
\caption{\label{fig:sp4} As in Figure \ref{fig:sp1} for PSR \pulsareight~(left panel)~and PSR \pulsarnine~(right panel). Colors of the labels in the upper panels denote the corresponding data in Chandra, {\em XMM-Newton}, Swift and {\em Suzaku}. The Chandra, {\em XMM-Newton}, Swift and {\em Suzaku} data are shown by different colors.}
\end{figure*}

We grouped the resulting individual spectra to have at least 25 counts per channel. For pulsars PSR~\pulsarthree~and PSR~\pulsarone~ for which we have more than one spectrum, we fit all the extracted spectra simultaneously. For all the fits we used XSPEC\footnote{\url{https://heasarc.nasa.gov/docs/xanadu/xspec/index.html}} version 12.9.0 \cite{Arnaud1996}. We took into account the effect of the interstellar absorption, using the {\em tbabs} model \cite{Wilms&McCray2000} assuming interstellar abundances for each source. We fixed the value of the hydrogen column density to the weighted average of the value given by \cite{Kalberla2005} in the direction of the sources in Table \ref{t_sp_res}, using the FTOOL {\it nH}
\footnote{\url{https://heasarc.gsfc.nasa.gov/lheasoft/ftools/heasarc.html}}. 
Below we discuss the effects of this assumption on our results.
Thereafter, we used a thermal~({\em Blackbody}) and a non-thermal~({\em Power-law}) component to model the observed spectra of each source. Note that in some cases only a {\em Blackbody}~(BB) or a {\em Power-law} model already yielded a statistically acceptable fit, rendering the introduction of a second component unnecessary. In these cases we used only one model component. For instance, the obtained X-ray spectra of PSR~\pulsarone~required only a non-thermal component, while PSR~\pulsarthree~data requires a two component model. The results of these fits are given in Table \ref{t_sp_res} and the spectra together with the best fit models are shown in Figures \ref{fig:sp1}, \ref{fig:sp2}, \ref{fig:sp3}, and \ref{fig:sp4}. 
For each pulsar unabsorbed fluxes are calculated for the 0.2-10 keV energy range and given in units of $10^{-14}$~erg/s/cm$^{-2}$ in Table \ref{t_sp_res}. 
The flux values correspond only to the thermal component whose temperature is also given in Table \ref{t_sp_res}, if available. Otherwise, it represents the total flux.

We used two X-ray observations of PSR~\pulsarone, one obtained on 29th June 2011 by CXO and another most recent one with {\em XMM-Newton} on 16th October 2017. These two observations show that, while the X-ray spectral shape does not change and can be fitted with a simple absorbed {\em Power-law}, the flux changes by a factor of 3. The flux we measure with CXO is $F_{pow} = 2.75_{-0.06}^{+0.08}\times10^{-14}~{\rm {erg}\,{\rm {cm}^{-2}\,{\rm {s}^{-1}}}}$ and the flux of the source measured by {\em XMM-Newton} is found to be  $F_{pow} = (8.55\pm{0.77})\times10^{-14}~{\rm {erg}\,{\rm {cm}^{-2}\,{\rm {s}^{-1}}}}$.

In our analysis of PSR~\pulsarfive, we found two different solutions each requiring a thermal component. The role of these components however seem different. We can fit the X-ray spectra with a {\em blackbody} plus a {\em Power-law} component and find the best fit parameters as, $kT_{BB}$ = $120\pm{11}$~eV and $\Gamma$ = $1.15\pm{0.018}$ for a fixed hydrogen column density as $0.37\times10^{22}~cm^{-2}$, with $\chi^2_{\nu}$ / dof = 0.902/417. Utilizing the precise distance of this source, the normalization of the {\em blackbody} component inferred from the fit corresponds to an apparent emitting radius of $\sim0.88$~km, indicating the existence of a hot spot on the surface. However the same data can also be fit with a neutron star atmosphere model plus a {\em Power-law}, where the parameters can be set to reflect the full surface component, with a marginally better fit. For the same hydrogen column density, assuming a zero magnetic field and a fixed neutron star mass at 1.4~M$_{\odot}$ for three different assumed radii of the neutron star at the distance given in Table \ref{t_msp_lit} we infer a surface temperature of $\rm{kT}$ = $46.8_{-7.1}^{+8.9} \rm{eV}$ with a similar {\em Power-law} slope and a $\chi^2_{\nu}$ / dof = 0.89/418. The difference in the degrees of freedom here comes from the fact that in the NSA fit the distance and the radius is also fixed at a value and only the surface temperature is the free fit parameter. Archival X-ray observations of PSR~\pulsarfive~ are not suitable for a pulse phase resolved spectroscopy due to their time resolution, therefore it is impossible to discriminate between the hot spot or full surface solutions using existing X-ray data. A future observation with {\em NICER} may help resolve this issue, thanks to {\em NICER}'s high time resolution and large effective area in the soft X-rays \cite{Gendreau2017}. We assume the temperature value obtained from the NSA modeling is the surface temperature of the neutron star and use this value as our surface temperature. Still for comparison we also show the limit on the {\em blackbody} temperature from the whole surface.


\begin{table*}
\centering \caption{Exposure times of the observations of MSPs analyzed in this paper and the net counts detected from the sources within the studied energy ranges.}
\begin{tabular}{cccccc}
\hline 
\hline 
 &\multicolumn{3}{c}{Total Exposure Time (ks)} \tabularnewline
Source Name  & Chandra & XMM-{\it Newton} & SUZAKU & Total Net Counts 
\tabularnewline
\hline
\noindent {\em PSR~\pulsarone}&$20.71$&$82.0$&$-$&$594$&\\ 
\noindent {\em PSR~\pulsartwo}&$64.14$&$-$&$-$&$315$&\\ 
\noindent {\em PSR~\pulsarthree}&$10.07$&$29.82$&$78.95$&$7792$&\\ 
\noindent {\em PSR~\pulsarfour}&$20.05$&$-$&$-$&$142$&\\ 
\noindent {\em PSR~\pulsarfive}&$55.07$&$62.08$&$-$&$17562$&\\ 
\noindent {\em PSR~\pulsarsix}&$-$&$40.9$&$-$&$316$&\\ 
\noindent {\em PSR~\pulsareight}&$-$&$71.2$&$-$&$219$&\\ 
\noindent {\em PSR~\pulsarnine}&$-$&$51.99$&$-$&$1394$&\\ 
\hline 
\end{tabular}
\label{t_obs}\\
\begin{raggedright}
\end{raggedright}
\end{table*}
\begin{table*}
\centering \caption{Observational properties of the millisecond pulsars as compiled from the literature.}
\begin{tabular}{cccccccccccc}
\hline 
\hline 
Source Name  & Type$^{1}$ & $f$ & $-\dot{f}$&Distance$^{2}$& Ref. \tabularnewline
 &  & (Hz) & ($10^{-16}\,{\rm s^{-2}}$)&(kpc)& \tabularnewline
\hline
\noindent {\em \pulsarone}&BW&$602.41$&$-$&$2.0$&$1$ \tabularnewline 

\noindent {\em \pulsartwo}&BW&$245.43$&$5.38$&$0.410\pm{0.008}$&$2$ \tabularnewline 

\noindent {\em \pulsarthree}&R&$271.45$& $16.8$& $0.44$&$3$ \tabularnewline
 
\noindent {\em \pulsarfour}&BW&$436.68$&$-$ & $0.6$&$4$ \tabularnewline 
\noindent {}&&&&$1.33^3$& \tabularnewline
\noindent {\em \pulsarfive}&R&$538.87$&$\sim7.5\times10^{-21}$&$0.91_{-0.04}^{+0.05}$&$5$ \tabularnewline 
\noindent {\em \pulsarsix}&$-$&$324.23$&$7.60$& $0.27$&$6$ \tabularnewline 
\noindent {\em \pulsareight}&BW&$707.29$&$4.6\times~10^{-21}$& $1.74$&$7$ \tabularnewline 
\noindent {}&&&&$0.97$&$8$ \tabularnewline
\noindent {\em \pulsarnine}&BW&$457.31$&$13.88$& $0.49$&$9$ \tabularnewline
\noindent {}&&&&$0.96^3$& \tabularnewline
\hline 
\end{tabular}\\
{\footnotesize{}{ }{\footnotesize \par}
\begin{raggedright}
{\footnotesize{}$^{1}$ BW : Black-Widow, R : Redback }\\
{\footnotesize{}$^{2}$For the distance based on the dispersion-measure the NE2001 model is used \cite{CordesLazio2002}.\\
{\footnotesize{}$^{3}$Distance value that is derived using the YMW16 model (\cite{Yao2017}) for the Galactic distribution of free electrons}.}\\ 
\par\end{raggedright}{\footnotesize \par}
{\footnotesize{} References: (1) \cite{Gentile2014}; \cite{Breton2013}; \cite{Polzin2018} (2) \cite{Toscano1999}; \cite{Perera2019} (3) \cite{Schwenzer}; \cite{Bassa2016} (4) \cite{Gentile2014} (5) \cite{Jennings2018}; \cite{Kong2017}; \cite{vanStaden2016} (6) \cite{Swiggum2017}; (7) and (8) \cite{Bassa2017}, \cite{Ho:2019myl}; (9) \cite{Keth2011}}.\\
}{\footnotesize \par}
\label{t_msp_lit}
\end{table*}

\begin{table*}
\centering \caption{Results of the spectral analysis of all sources.}
\begin{tabular}{cccccccccc}
\hline
\hline
Source Name&$nH^*$&$kT$&R$^\dagger$&$\Gamma$&Flux& $\chi^2_{\nu}$ / dof \tabularnewline
& (10$^{22})$ cm$^{-2}$ &(keV)&(km)&&$(10^{-14}$erg/s/cm$^{-2})$ & \tabularnewline
\hline 
\noindent {\em PSR~\pulsarone}&$0.09$&$-$&$-$&$1.72\pm{0.09}$&$8.55\pm0.77$&0.8/15\\
\noindent {\em PSR~\pulsartwo}&$0.21$&$-$&$-$&$2.57 \pm{0.21}$&$7.37\pm0.85$&1.09/16\\

\noindent {\em PSR~\pulsarthree}&$0.04$&$0.16\pm0.01$&$0.159_{-0.041}^{+0.144}$&$2.85\pm{0.05}$&$30.80\pm1.20$&0.98/225 \\
\noindent {\em PSR~\pulsarfour}&$0.03$&$-$&$-$&$2.93\pm0.20$&$6.46\pm1.30$&0.73/6\\
\noindent {\em PSR~\pulsarfive}&$0.37$&$0.12\pm{0.11}$&$0.88\pm{0.20}$&$1.15\pm{0.018}$&$177.05\pm2.7$&0.89/418\\
\noindent {\em PSR~\pulsarsix}&$0.02^{**}$&$0.16\pm{0.01}$&$0.04\pm{0.01}$&$-$&$1.16\pm0.17$&0.51/18\\
\noindent {\em PSR~\pulsareight}&$0.04$&$-$&$-$&$2.44_{-0.21}^{+0.23}$&$1.07\pm0.04$&1.21/12\\
\noindent {\em PSR~\pulsarnine}&$0.01$&$0.176\pm{0.01}$&$0.06\pm0.02$&$2.64\pm0.15$&$6.23\pm{0.02}$&0.56/25\\
\hline 
\end{tabular}
\label{t_sp_res}\\
\begin{raggedright} 
{\footnotesize{}$*$ {\em nH} values have been obtained from \cite{Kalberla2005}.\\ 
$^{**}$ The value for this source has been taken from \cite{Swiggum2017}.\\ 
$^\dagger$ R is the apparent emitting radius calculated using the distance and the normalization of the blackbody model.}
\end{raggedright}
\end{table*} 
\begin{table*}
\centering \caption{Results of the NSA and blackbody temperature for different radius and bounds on the surface temperature in the considered spectral model. The error range for the temperatures stems from the uncertainties of both the radius and  distance measurements. The radius is considered to be in the range $8 {\rm ~km} \leq  R \leq 15 {\rm ~km}$ and the uncertainty in the distance is assumed to be  $\pm$15\% in cases where it is not provided. The largest temperature corresponds to the smallest radius and largest source distance, and vice versa.}

\begin{tabular}{ccccc|cccc} 
\hline
\hline 
Source name&$f$&Distance&$T_\infty^{({\rm BB})}$ & $T_s^{({\rm NSA})}$&$\alpha_{\rm fiducial}^{\rm (NSA)}$& $\alpha_{\rm rigorous}^{\rm (NSA)}$& \\
             & (Hz)  & (kpc)        & (eV)         & (eV)  & & & \\
\hline
\noindent {\em \pulsarone}&$602.41$&$2.00$&$37.0_{-2.0}^{+4.0}$&$23.6_{-8.9}^{+5.1}$ & $1.7\times 10^{-9}$&$4.0\times 10^{-9}$& \\
\noindent {\em \pulsartwo}&$245.43$&$0.410\pm{0.008}$&$33.0_{-2.0}^{+1.0}$&$24.5_{-3.5}^{+4.8}$&$6.5\times 10^{-8}$& $1.5\times 10^{-7}$& \\  
\noindent {\em \pulsarthree}&$271.45$&$0.44$&$27.0_{-2.0}^{+3.0}$&$26.2_{-11.0}^{+8.8}$& $5.0\times 10^{-8}$& $1.4\times 10^{-7}$&\\ 
\noindent {\em \pulsarfour*}&$436.68$&$0.59$&$40.0_{-2.0}^{+3.0}$&$23.3_{-6.0}^{+3.3}$&$1.2\times 10^{-8}$&$3.0\times 10^{-8}$&\\
\noindent {}&$436.68$&$1.33$&$46.0_{-2.0}^{+4.0}$&$33.01_{-7.89}^{+8.06}$&$-$&$-$&\\
\noindent {\em \pulsarfive**}&$538.87$&$0.91_{-0.04}^{+0.05}$&$57.80_{-3.3}^{+5.2}$&$46.8_{-7.1}^{+8.9}$& $1.0\times 10^{-8}$& $2.3\times 10^{-8}$\\ 
\noindent {\em \pulsarsix}&$324.23$&$0.27$&$17.0\pm{1.0}$&$10.5_{-1.7}^{+1.5}$& $3.9\times 10^{-9}$& $8.3\times 10^{-9}$&\\
\noindent {\em \pulsareight*}&$707.29$&$1.74$&$34.0_{-3.0}^{+4.0}$&$28.07_{-5.96}^{+7.03}$& $1.2\times 10^{-9}$& $3.1\times 10^{-9}$&\\

\noindent {}&$707.29$&$0.97$&$17.0\pm{3.0}$&$22.66_{-3.08}^{+3.13}$&\\
\noindent {\em \pulsarnine*}&$457.31$&$0.49$&$31.0\pm{3.0}$&$21.7_{-5.6}^{+4.2}$&$7.1\times 10^{-9}$&$1.7\times 10^{-8}$&\\
\noindent {}&$457.31$&$0.96$&$36.0_{-3.0}^{+2.0}$&$28.04_{-7.17}^{+6.10}$&$-$&$-$&\\
\hline
\end{tabular}
\label{t_temp_lim}\\
\begin{raggedright}
{\footnotesize{}$^*$} Two different distance values ($d=0.59\,{\rm kpc}$ - $ d= 1.33\,{\rm kpc}$, $d=1.74\,{\rm kpc}$ - $ d= 0.97\,{\rm kpc}$ and $d=0.49\,{\rm kpc}$ - $ d= 0.96\,{\rm kpc}$, respectively) are assumed.\\
{\footnotesize{}$^{**}$} Note that for \pulsarfive ~the $T_\infty^{({\rm BB})}$ value is a limit on the temperature, whereas the $T_s^{({\rm NSA})}$ is the measured temperature as detailed in Section~\ref{sec:analysis}. \\
\end{raggedright}
\end{table*}
\begin{figure*}
\includegraphics[scale=0.40]{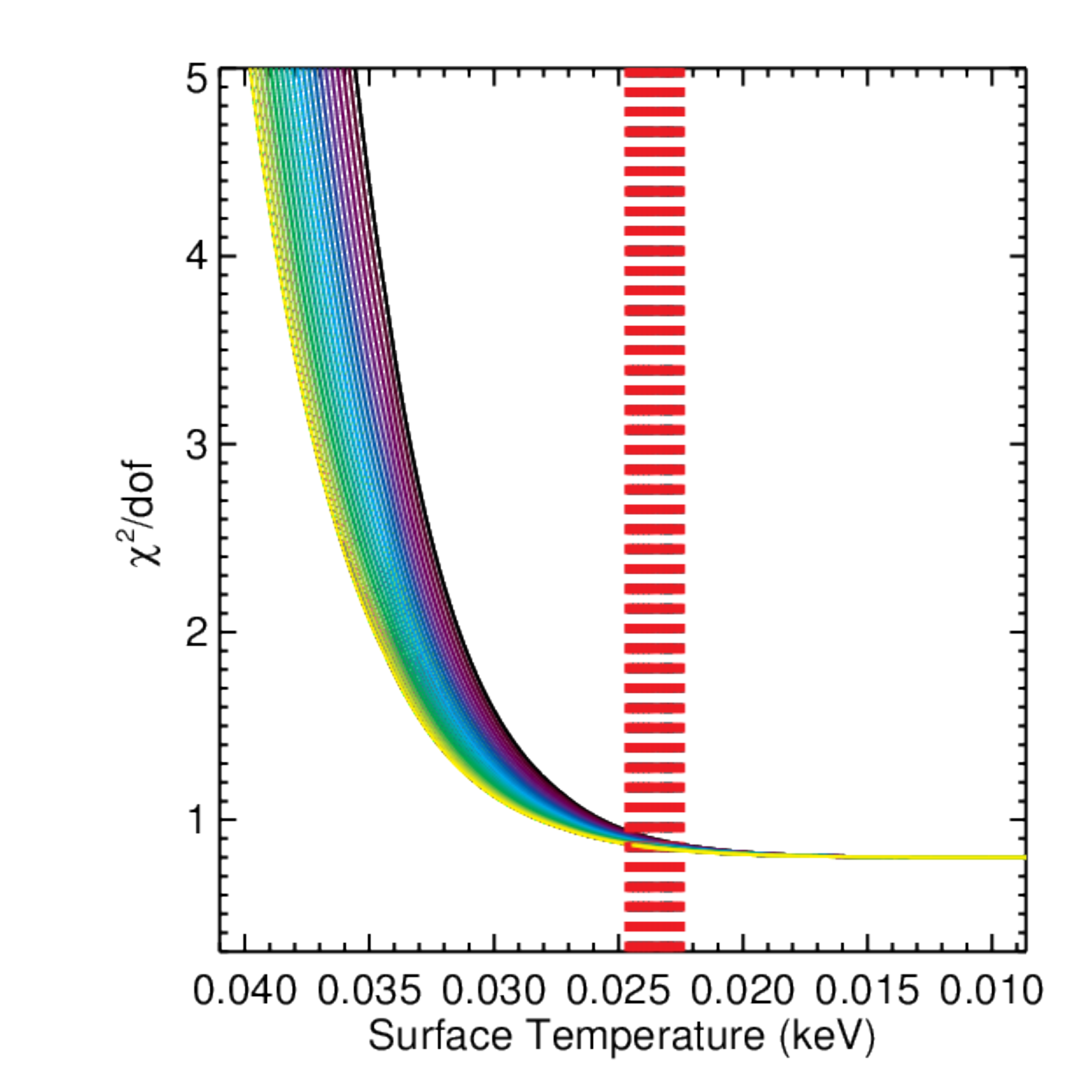}
\caption{\label{fig:tlim1} $\chi^2/dof$ as a function of the unredshifted surface temperature of the assumed neutron star atmosphere model for PSR \pulsarone. The limiting temperature is determined from the point where the $\chi^2/dof$ deviates from the best fit value by 1$\sigma$ (Avni et al., 1976). The thickness of the vertical shaded area is due to different distance and radius assumptions.}
\end{figure*}

\begin{figure*}
\includegraphics[scale=0.40]{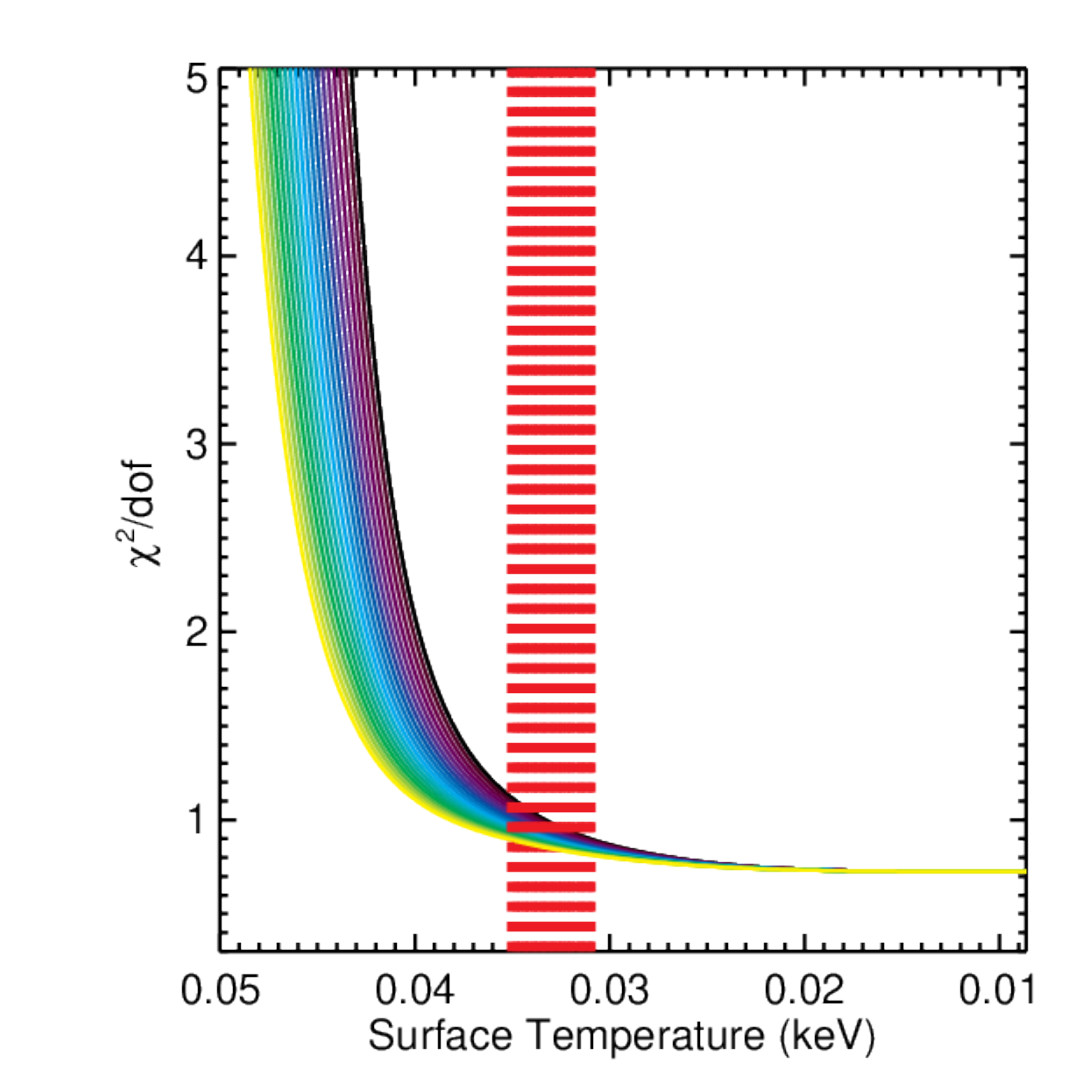}
\includegraphics[scale=0.40]{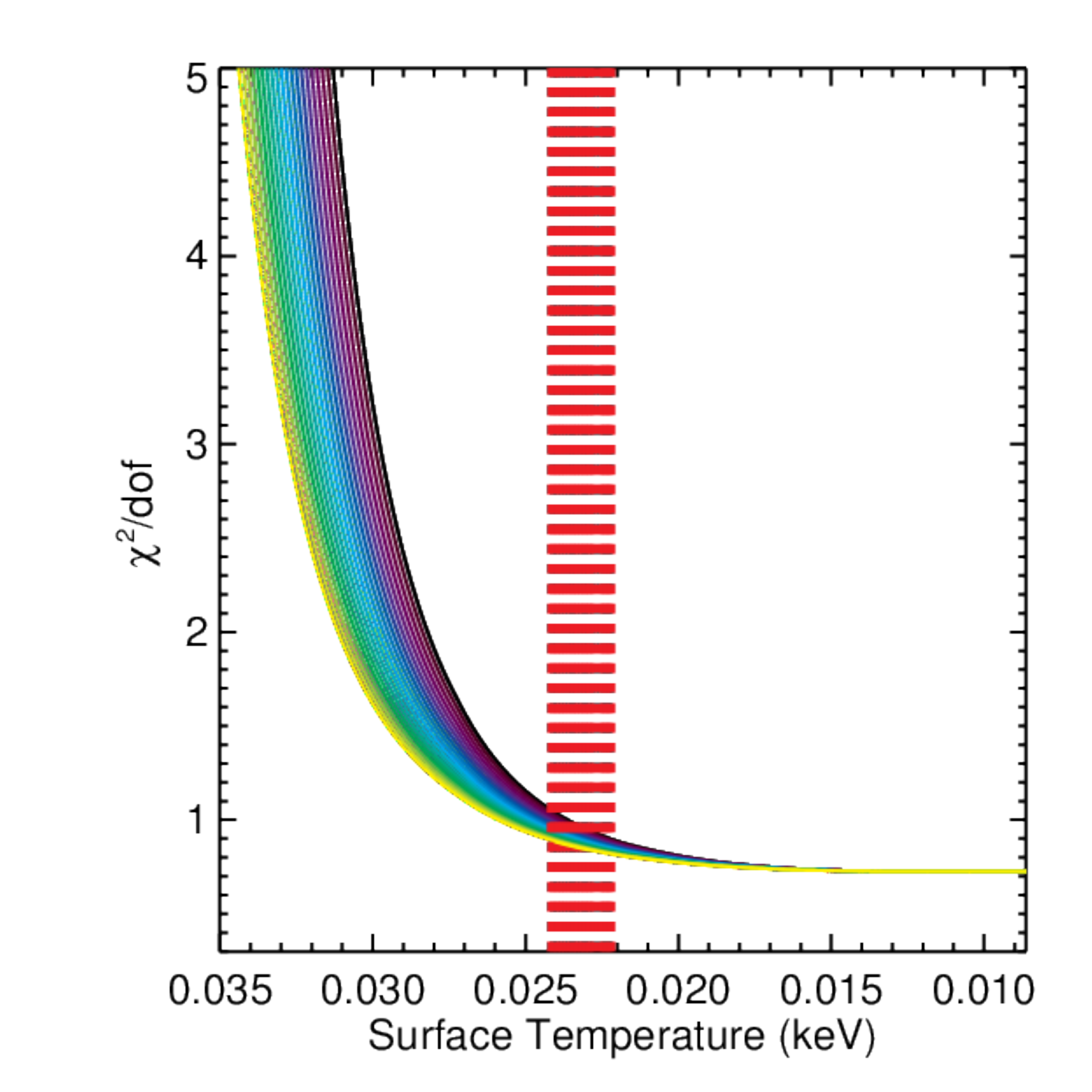}
\caption{\label{fig:tlim2} As in Figure \ref{fig:tlim1}, for PSR \pulsarfour, there are two assumed distances, 1.33~kpc~(left panel) and 0.59~kpc~(right panel).}
\end{figure*}


\begin{figure*}
\includegraphics[scale=0.40]{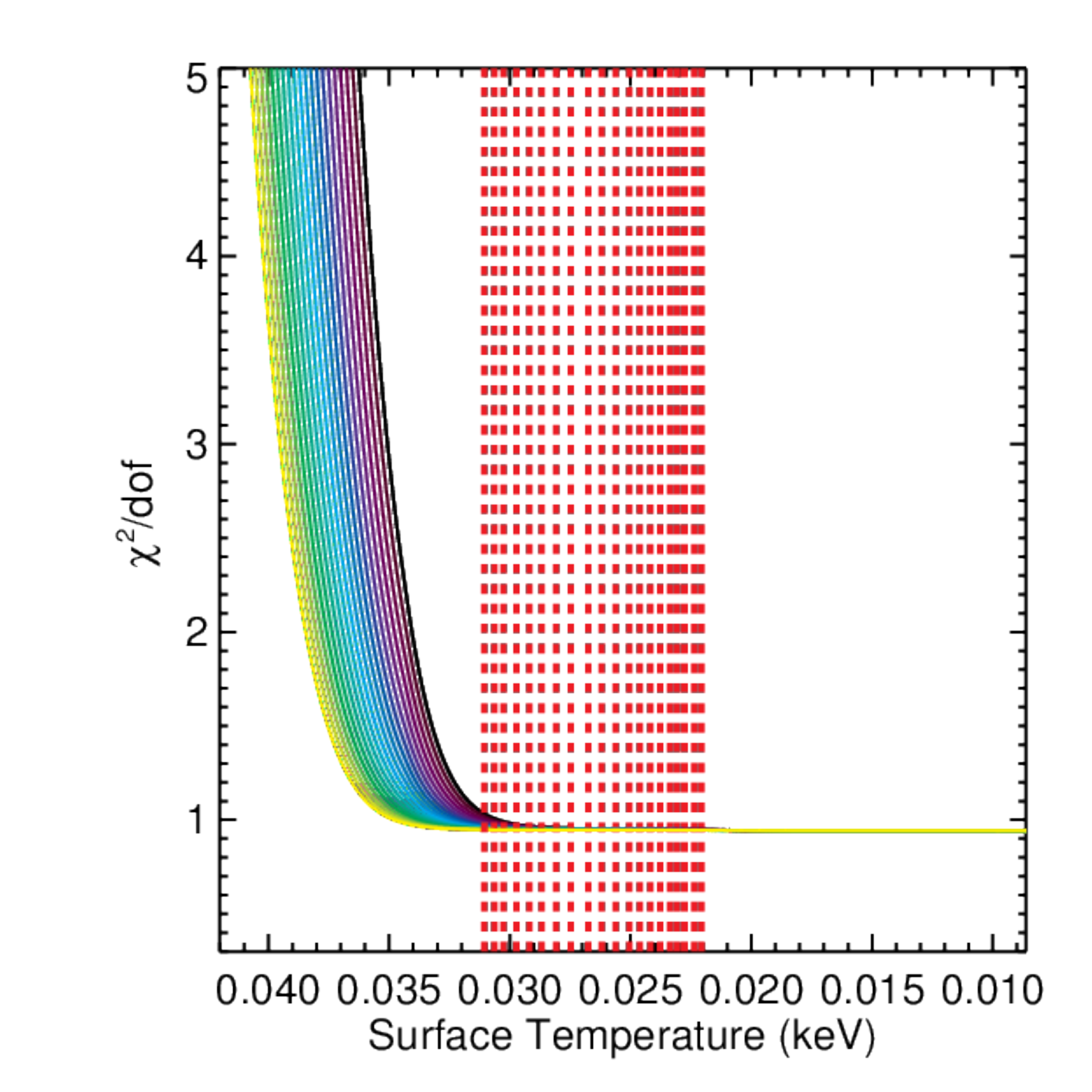}
\includegraphics[scale=0.40]{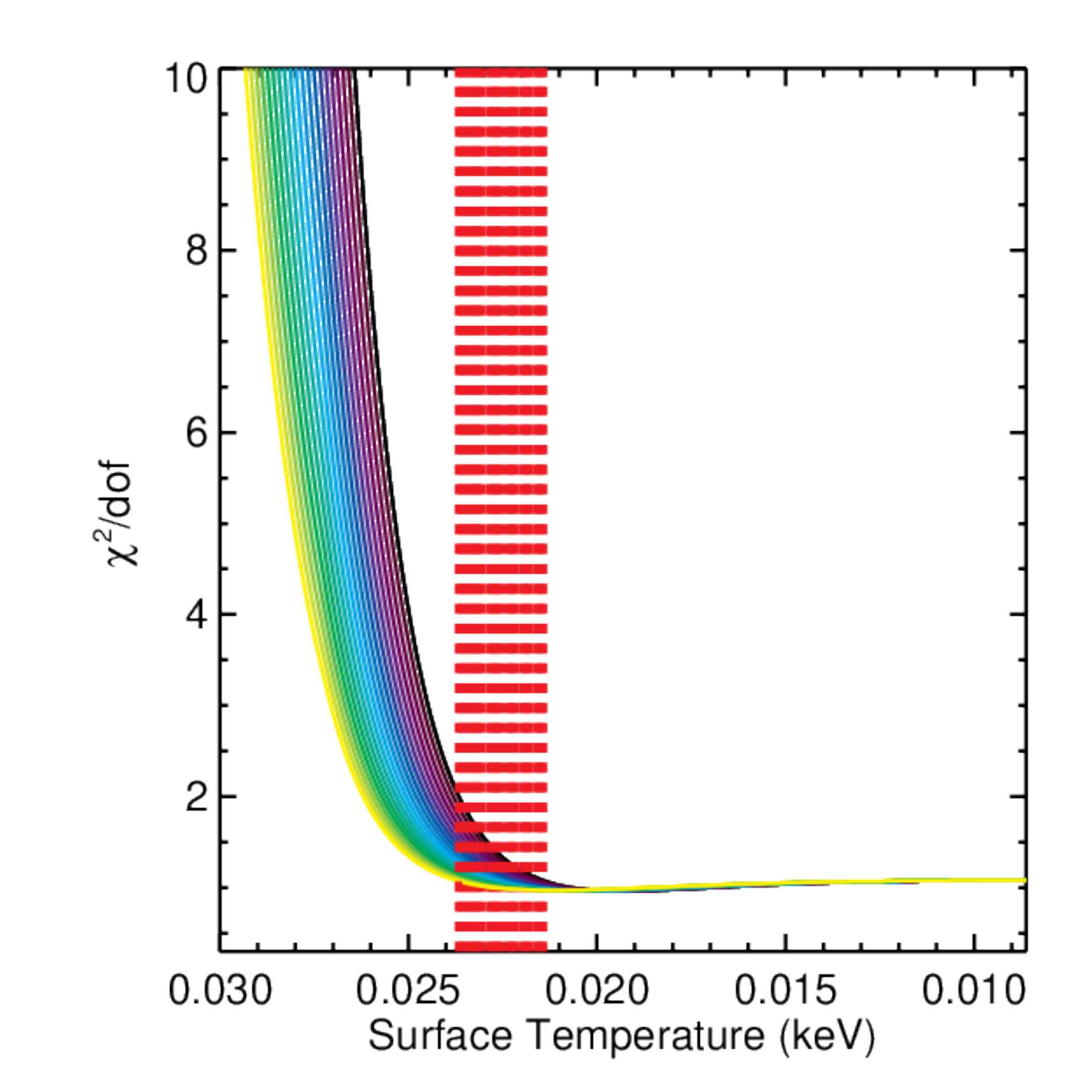}
\caption{\label{fig:tlim3} As in Figure \ref{fig:tlim1}, results for PSR \pulsarthree~(left panel)~and PSR \pulsareight. Note that for PSR \pulsareight~the limit on the surface temperature does not show a significant variation despite different distance assumption (right panel). 
 }
\end{figure*}

\begin{figure*}
\includegraphics[scale=0.40]{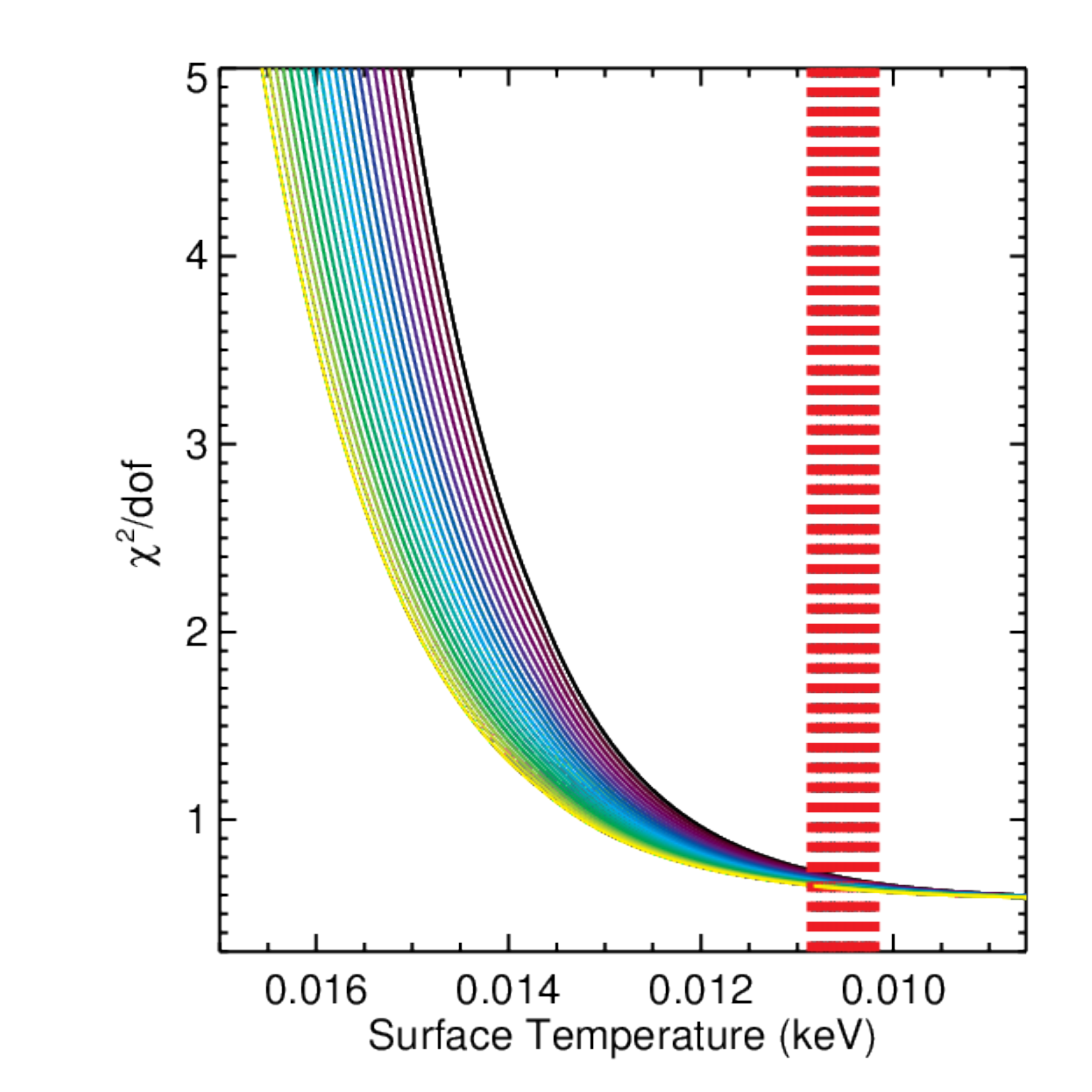}
\includegraphics[scale=0.40]{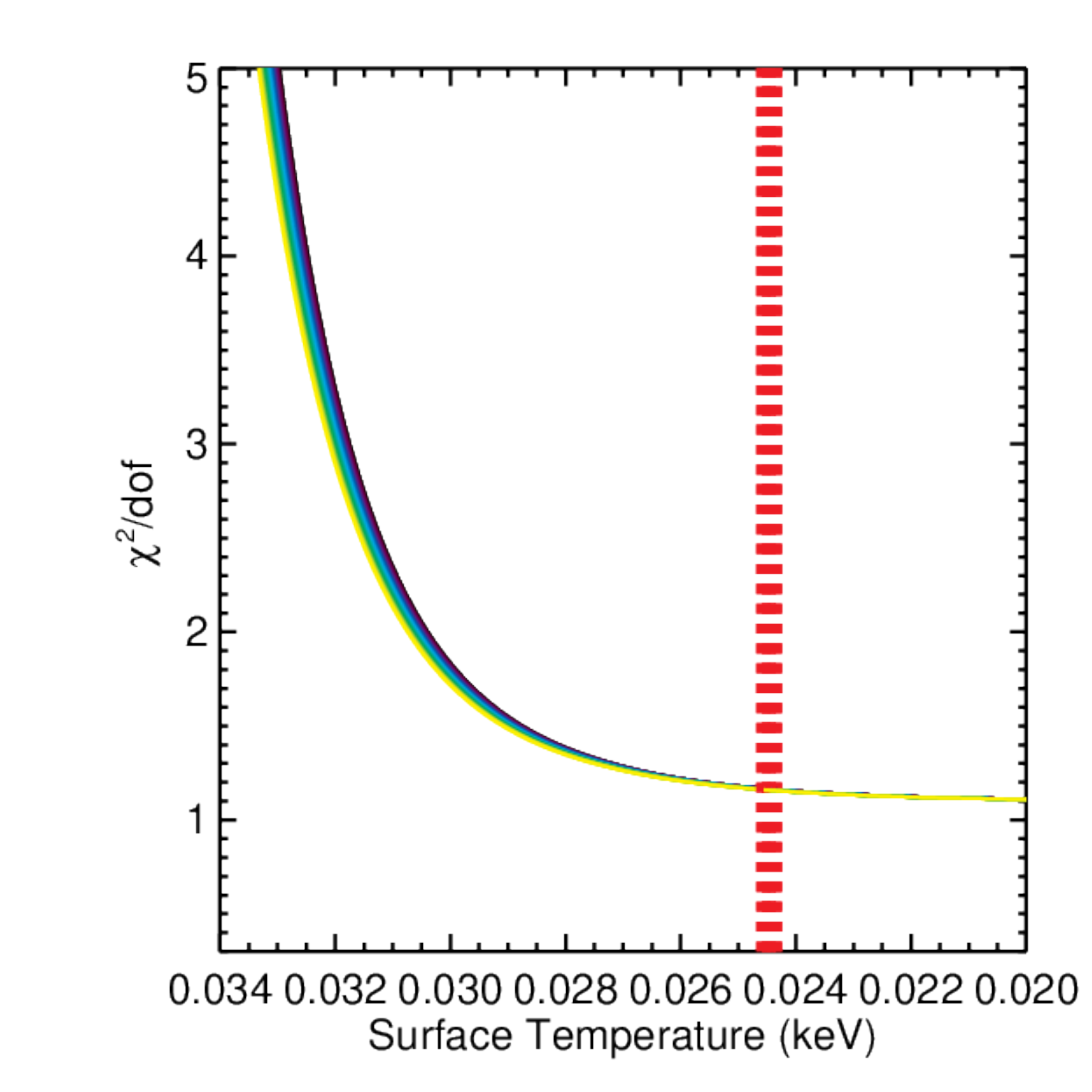}
\caption{\label{fig:tlim4} As in Figure \ref{fig:tlim1}, for PSR \pulsarsix~(left panel) and PSR \pulsartwo~(right panel).} 
\end{figure*}

\begin{figure*}
\includegraphics[scale=0.40]{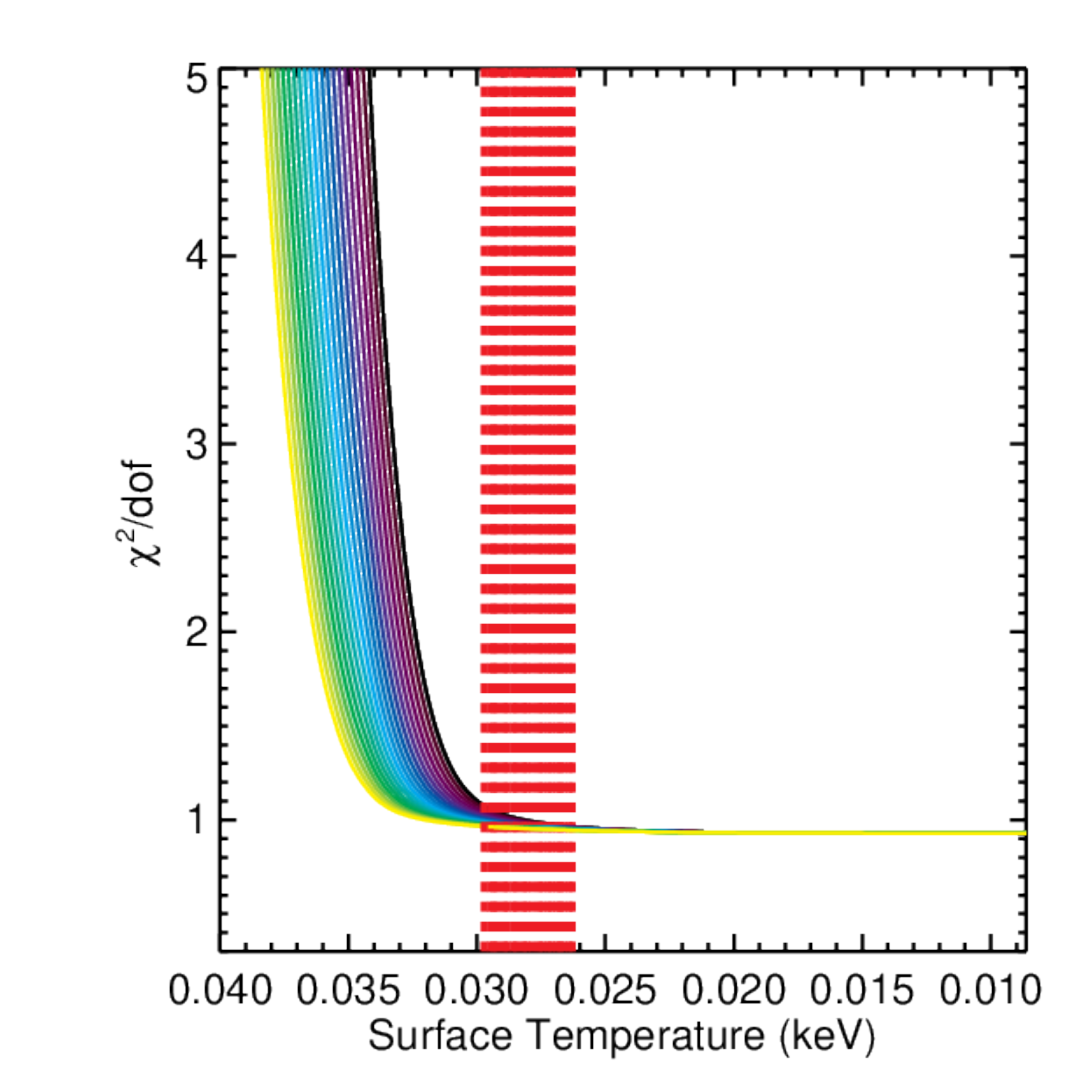}
\includegraphics[scale=0.40]{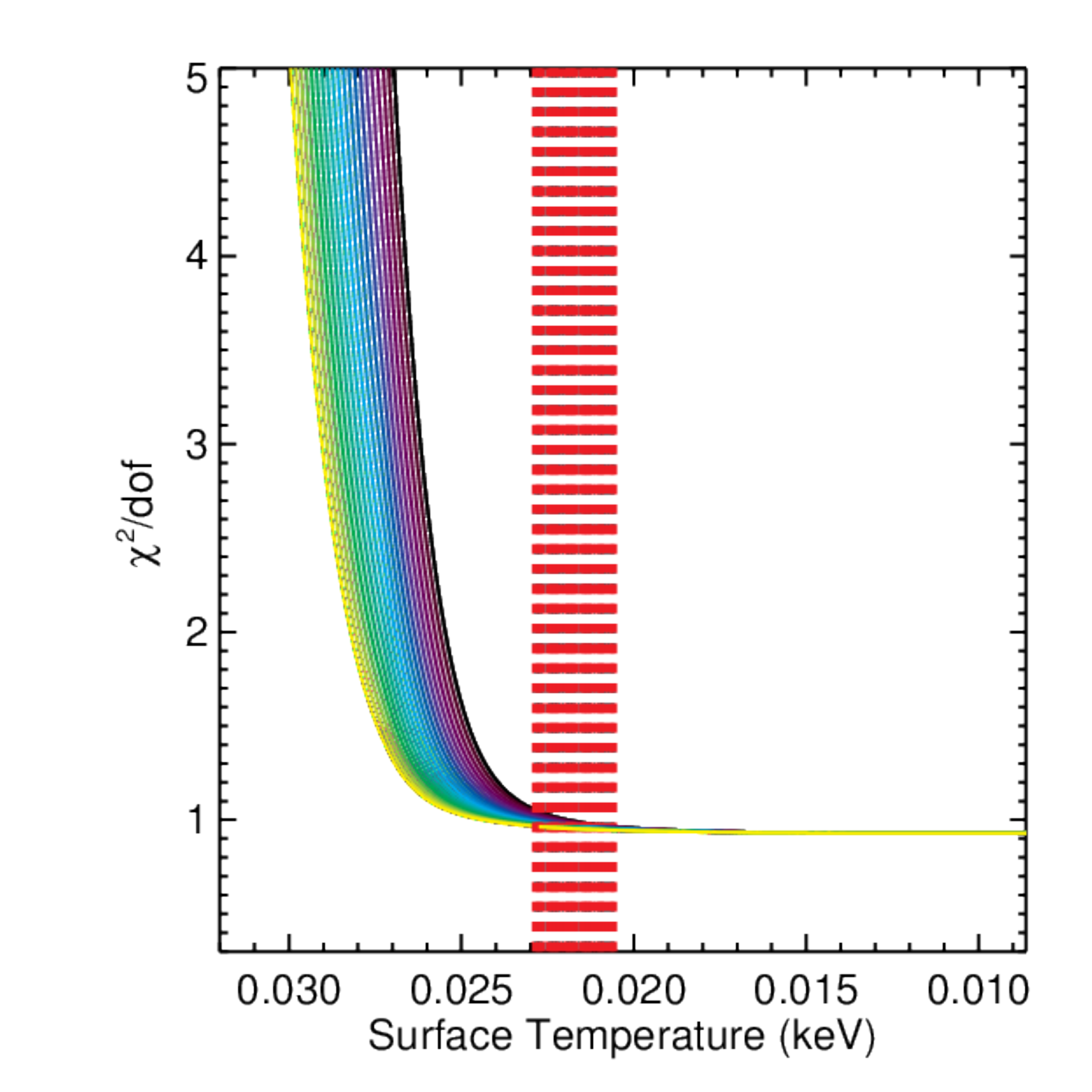}
\caption{\label{fig:tlim5} As in Figure \ref{fig:tlim1}, temperature limit for PSR \pulsarnine. Results for two assumed distances, 0.96kpc~(left panel) and 0.49kpc~(right panel), are shown.
 }
\end{figure*}


\section{Limits on Thermal Emission from X-ray data}\label{sec:limits}

Once the modeling of the observed spectra was completed, we moved on to apply the method outlined in \cite{Schwenzer}, which is to find the highest temperature value of a thermal component emitted from the whole surface that would cause a statistically significant deviation on the existing model, for an assumed neutron star mass, radius and distance. In this paper, we make several improvements to our assumptions in the previous paper \cite{Schwenzer}.

First we improve on our previous assumption that the thermal component from the surface can be represented with a simple BB emission. In addition to a blackbody, we here also employ a fully ionized Hydrogen atmosphere model, namely {\em NSA} in Xspec \citep{Zavlin1996}, in radiative and hydrostatic equilibrium. It is well known that even if there is only a small amount of matter collected on the surface of a neutron star, such a layer can act as an atmosphere, and it significantly changes the energy distribution of the photons emanating from the surface \citep{Zavlin1996, ozel2013}. 
Typically, such an atmosphere results in broader spectra compared to a pure BB emission, because of the energy dependence of the opacity in the atmosphere. Due to this dependence, more energetic photons at deeper layers effectively become visible to an observer, resulting in more flux at higher energies and hence a broader observed X-ray spectrum. If an observed spectrum from such an atmosphere is modeled with a simple BB function, then to provide a fit to the data with its narrow spectral shape, the best fit BB temperature and apparent radius values become higher and lower, respectively, compared with those obtained with a more realistic spectral model. This is why a temperature obtained by fitting an atmospheric emission with a {\em blackbody} is called {\em color temperature}. Using a more realistic spectral model, like NSA, therefore naturally results in a smaller temperature and a larger emitting radius. The fact that NSA or similar atmosphere model spectra are broader than a BB in particular significantly helps to limit the faint surface emission, since they can better reproduce the high energy tail of the atmosphere. Therefore, they increase the sensitivity of standard X-ray spectroscopy to such low-energy signals, which likely peak in the UV: As we increase the predicted temperature of the atmosphere to higher values, to see if it affects our fit to the observed spectra, a broader spectrum obviously has a statistically more significant effect on the fit than a narrower (pure {\em blackbody}) model \cite{ozel2013}. Note that for comparison, we also provide limits on the surface temperature of these sources assuming a BB function.

The {\em NSA} model provides tabulated X-ray  spectra in the 0.05 to 10~keV range. There are three options for the strength of the magnetic field: B = $0$, $10^{12}$, $10^{13}$~G. Because the {\em NSA} model assumes a fully ionized Hydrogen atmosphere with Thomson scattering, it is only valid within the temperature range $10^5-10^7$~K \citep{Zavlin1996}. The normalization of the model is defined as the inverse square of the distance of the object in parsecs. This model takes into account the gravitational redshift and so the best-fit resulting temperatures are unredshifted temperatures $T_s$ at the surface of a neutron star for an assumed mass and radius. For the purpose of this paper we used $0$~G magnetic field strength, as is the most appropriate case for millisecond pulsars (having fields $\ll 10^{12}$~G). For the gravitational redshift, we took into account a fixed neutron star mass of $M_{\odot}$=1.4, and three radius values  $R$= 8, 10, 15~km. 

The distance values we compiled from literature and use here are given in Table \ref{t_msp_lit} together with appropriate uncertainties. Typically these measurements are derived from the inferred dispersion measurements and the Galactic electron density maps \citep{Weisberg1996, CordesLazio2002, Yao2017}. Especially for the distance measurements relying on electron density maps, often the uncertainties are not presented, since the model-dependent systematic uncertainties are often larger than the formal uncertainties of the measurements. In such cases we assumed a 15\% uncertainty on the distance and used this to derive our limits on the surface temperature. We note that the distances used here are often inferred from the NE2001 model \cite{CordesLazio2002}. Newer models like \cite{Yao2017} can provide distances with smaller uncertainties on average. For this reason, we compared the distances of all the sources we use with these two models. The distance values obtained are often very similar for most of the source, except for PSR~\pulsarfour~and PSR~\pulsarnine. Since the calculated distance values from different models can have a significant effect on our temperature limits, we present our results for two different distances in Table \ref{t_temp_lim}.

In addition to these, with the second data release of the GAIA\footnote{\url{http://sci.esa.int/gaia/}} mission (\cite{Lindegren2018}) astrometric parameters and distances of several millisecond pulsars have been calculated by \cite{Jennings2018}. In our analysis we have been able to use such a new measurement for PSR~\pulsarfive~. Moreover, for PSR \pulsartwo the distance has been calculated by radio parallax \cite{Perera2019}, giving a more precise value.


Taking into account the above mentioned atmospheric effects and uncertainties in the distance, we followed a method very similar to that used in \cite{Schwenzer}. We added a thermal component representing the surface emission of the neutron star to the best fit model, presented in Section~\ref{sec:analysis}. 
For a fixed distance and radius (apparent emitting radius in the case where we assumed a BB function for the surface emission), we increased the temperature of the thermal component from the lower limit of the NSA model (0.001~keV in the case of a BB model) and investigated the resulting $\chi^2/dof$ while allowing other fit parameters  presented in Table \ref{t_sp_res} to vary. Note that this is actually very similar to how the \emph{error} command on Xspec works but instead of defining confidence region of a best fit model parameter, here we are only finding the limit for an additional model component which is not needed to fit the data to begin with.
For each source the resulting change in the $\chi^2/dof$ as a function of the temperature of this hot surface component is shown in Figures \ref{fig:tlim1}, \ref{fig:tlim2}, \ref{fig:tlim3}, \ref{fig:tlim4} and \ref{fig:tlim5}. Colors in the figures indicate results for different distance and apparent radius assumptions. We report the limiting temperatures in Table \ref{t_temp_lim}, which are the values corresponding to a  1$\sigma$ in $\chi^2$ \cite{Avni1976}. 
The uncertainties in Table \ref{t_temp_lim} include the uncertainties in the distance and radius of the neutron stars. 

 One remaining assumption in our calculations is related to the Hydrogen column density. As it is well known, soft X-rays are absorbed in the interstellar medium (ISM) because of photoelectric absorption and scattering by gas and dust grains. For the analysis of thermal emission from neutron stars, generally, $\rm{N}_{\rm{H}}$ is a fit parameter that shows a correlation with the inferred temperature and emitting radius values. We here utilized the values presented by \cite{Kalberla2005} and used them as a frozen parameter in our analyses (see Table \ref{t_sp_res}). To test the potential effects of this assumption, we allowed the column density to be free while fitting the spectra as well as the other parameters. For this purpose we used four sources with thermal and non-thermal components. These sources are PSR \pulsarone, PSR \pulsarthree, PSR \pulsarfive~and PSR \pulsarnine~ respectively. 
 
We found that for three pulsars (PSR~\pulsarthree, PSR~\pulsarone ~and PSR~\pulsarnine) the inferred $\rm{N}_{\rm{H}}$ values increased by more than $49\%$. 
Such a change in the $\rm{N}_{\rm{H}}$ resulted in the limiting temperatures to change to $T_{\rm NSA}$ = $37.15$~eV, $T_{\rm NSA}$ = $25.50$~eV and $T_{\rm NSA}$ = $45.23$~eV, respectively. 
In one other case, we found the $\rm{N}_{\rm{H}}$ to be lower than the fixed value we used, by more than $35\%$, which resulted in a decrease in the limiting temperature, as $T_{\rm NSA}$ =$18.45$ ~eV, for PSR~\pulsarfive.
In several cases the hydrogen column density increases, when it is allowed to vary in the fits. However, note that the values we use are already integrated values along a given line of sight \citep{Kalberla2005} for the whole Galaxy, whereas the sources we are interested in are much closer. Therefore, it is unlikely that the increase we observe in the hydrogen column density values is real. 
Such variations indicate the complicated correlation between the inferred temperature and Hydrogen column density, which also depend on the signal to noise at lower energies.

Finally, as stated above we calculate the limits on the surface temperature by allowing the other parameters of the spectral model to vary. However it is also possible to fix the best fit parameters to values given in Table \ref{t_sp_res} and search for a change in the $\chi^2$ as a function of the assumed temperature of the thermal emission. Such an analysis results in temperature bounds mostly similar to the values presented here, but systematically lower, in a few cases (PSR~\pulsarthree~and PSR~\pulsareight) the change can be as much as 30\%. Note that in any case it is the data quality that determines the amount of such systematic uncertainties since it directly affects how well the spectral parameters are constrained. Future imaging observatories with large effective area in the soft X-rays may help resolve these issues and allow for better constraints, such as the  X-ray Recovery Imaging and Spectroscopy Mission (XRISM) (\cite{Ishisaki2018}) or the Advanced Telescope for High-Energy Astrophysics (Athena) (\cite{Nandra2013}, \cite{Pajot2018}).

\section{R-mode amplitude bounds}
\label{sec:rmode}
\hypersetup{draft}
As demonstrated previously (\cite{Mahmoodifar2013,Alford2013,Schwenzer}), the temperature bounds obtained in the last section impose bounds on the amplitude of r-modes (\cite{Papaloizou1978, Andersson1998, Lindblom1998, Friedman1998, Andersson2000}), which are global toroidal oscillations of rotating stars that are driven unstable by gravitational wave emission via the Friedman-Schutz mechanism (\cite{Friedman1978}).
Since r-modes are unstable, the dissipation required to saturate them would strongly heat the star if they are present. The observed rather cold millisecond sources therefore significantly constrain the presence of r-modes within them. The size of r-modes is determined by the dimensionless amplitude parameter $\alpha$ defined in \cite{Lindblom1998}.

Whereas spindown data was known to set bounds of at most $\alpha\lesssim10^{-7}$, X-ray data has steadily improved these limits. Initial limits stemmed from sources that were heated by accretion in LMXBs \cite{Mahmoodifar2013}, which allowed to directly measure their surface temperature, leading to bounds $\alpha\lesssim10^{-8}$. In \cite{Schwenzer} it was shown that even though the temperature of cold millisecond pulsars is too low to directly observe a thermal surface component, its absence can set even tighter constraints on the size of the r-mode amplitude, leading to bounds $\alpha\lesssim10^{-8}$ even when the sizable uncertainties in the analysis are taken into account. Similar results were presented for sources in globular clusters \cite{Bhattacharya2017} and very recently the spectral observation of the 707 Hz pulsar \pulsareight~set the most restrictive bound to date \cite{Ho:2019myl}.

Here we use the spectral results for the various millisecond pulsars discussed above, including in particular the novel data on PSR~\pulsarone, PSR~\pulsarnine~and the recent data on PSR~\pulsareight, to obtain tighter bounds on the r-mode amplitude. To this end, following \cite{Schwenzer}, we take into account the main uncertainties in the analysis to obtain robust upper bounds. As discussed in the previous section this includes in particular the use of the more realistic neutron star atmosphere (NSA) model.


The bound on the r-mode amplitude that is set by an observed temperature bound stems from simple conservation equations \cite{Mahmoodifar2013,Schwenzer} and has been given in its general form that exhibits the explicit dependence on the various parameters in \cite{Schwenzer}. The most important case are sources without fast neutrino cooling (like direct Urca processes), i.e. merely slow modified Urca processes, in which case for surface temperatures roughly below $10^{6}\,{\rm K}$, as realized for the sources considered here, photon cooling from the surface strongly dominates \cite{Schwenzer}. In contrast to a spectral blackbody model, the NSA model yields as an output the (unredshifted) surface temperature $T_s$ instead of the (redshifted) temperature $T_\infty$ as observed far away from the source. In terms of $T_s$ the bound on the r-mode amplitude for a star without fast neutrino cooling reads
\begin{equation}
\alpha_{{\rm sat}}\leq\sqrt{\frac{3^{7}5}{2^{25}\pi^{6}\left(3-2\chi\right)\chi^{5}G}}\frac{1}{\tilde{J}MR^{2}}\frac{T_s^{2}}{f^{4}}\label{eq:saturation-amplitude-bound}
\end{equation}
where $f$, $M$ and $R$ are spin frequency, mass and radius, while $\tilde{J}$ is a dimensionless constant entering the r-mode gravitational wave emission, that encodes the radial energy density profile $\rho (r)$ of the source
\begin{equation}
\tilde{J}_{m}\equiv\frac{1}{R^4 M}\int_{0}^{R}dr\,r^6\rho\left(r\right) >1/\left(20\pi\right) \label{eq:J-tilde-bound}
\end{equation}
and the rigorous lower bound is imposed by the stability of the star \cite{Alford:2014}. Since eq.~(\ref{eq:saturation-amplitude-bound}) relies on the temperature at the surface, in contrast to the corresponding expression in \cite{Schwenzer}, no red-shift factor arises\footnote{Note that in \cite{Schwenzer} there was unfortunately a typo in the definition of the red-shifted luminosity just before eq.~(5), which in this form should depend on the effective radius $R_\infty$, but it did not affect the bound in eq.~(5) and the results for the different sources in \cite{Schwenzer} or the present article.}.
The weakly frequency dependent factor $\chi\!\left(\Omega\right)\approx1$
describes the deviation of the connection between the rotation frequency
$f\!=\!\Omega/\left(2\pi\right)$ and the r-mode oscillation frequency
$\nu\!=\!\omega/\left(2\pi\right)$ from their canonical relation
$\omega\!=\,-\frac{4}{3}\chi\!(\Omega)\Omega$ and is determined by
general relativistic and rotation corrections.

\begin{figure}
\includegraphics[scale=0.65]{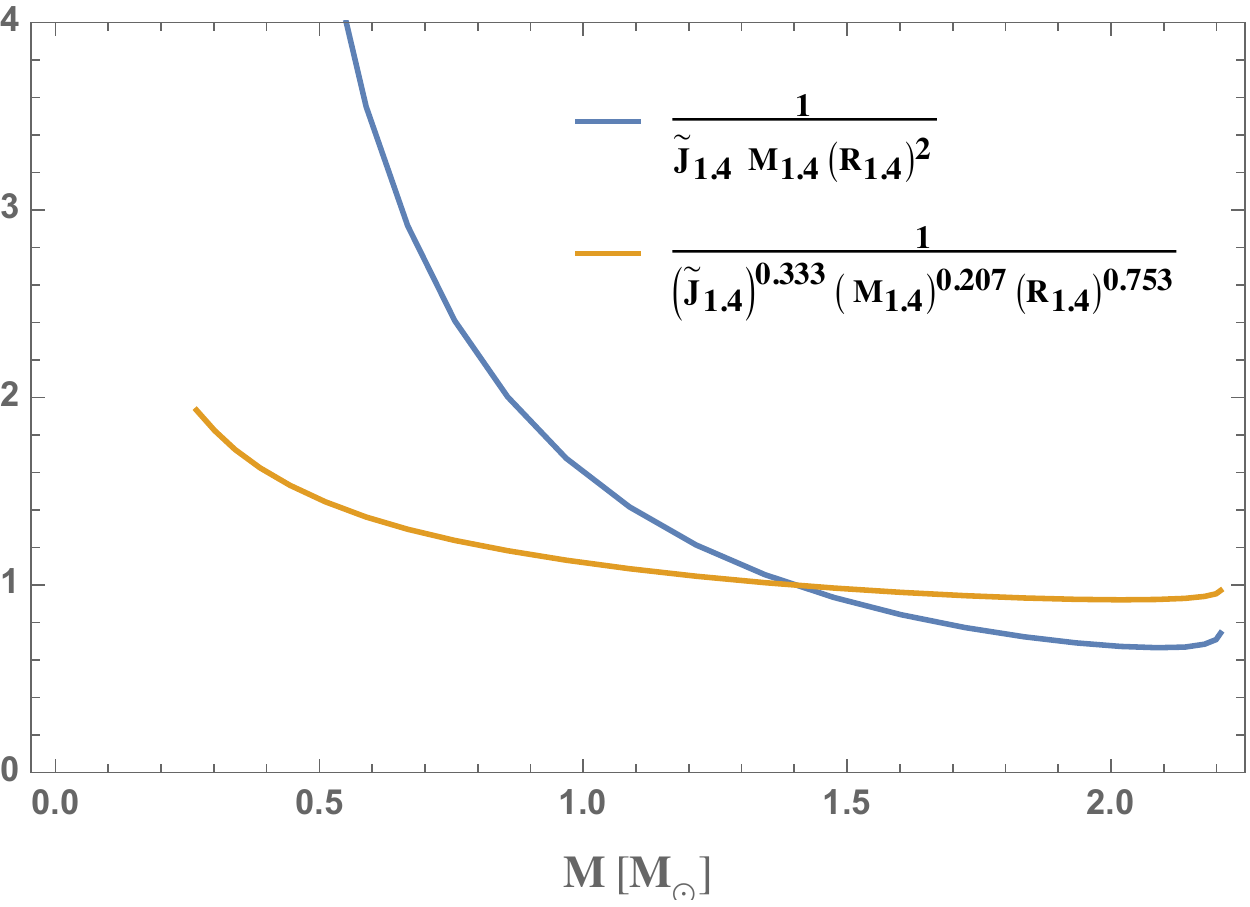}\caption{\label{fig:configuration-dependence}Factor that incorporates the dependence on the unknown source configuration in eqs.~(\ref{eq:saturation-amplitude-bound}) and (\ref{eq:frequency-bound}) as obtained from the solution of the TOV equation for an APR equation of state (Akmal et al. (1998)).}
\end{figure}

The various quantities arising in equation~(\ref{eq:saturation-amplitude-bound}) depend on the particular star configuration, which, for a given equation of state (EoS), can be parametrized by the mass $M$, and are not independent of each other. The radius in particular is approximately monotonously decreasing with increasing mass. In the non-relativistic case, valid for a neutron star sufficiently away from its mass limit, they scale roughly as $M \sim 1/R^3$. In this case the arising mass and radius dependence mostly cancels out and the expression would be merely linearly dependent on the radius.  Only close to the mass limit the mass asymptotes to its maximum value while the radius decreases further. Therefore, there the bound could become slightly weaker compared to the non-relativistic scaling region. Yet, this surely only happens when the bound is already particularly tight since the mass, which can vary over a large range, is at its upper limit. The same holds for $\tilde J$, which is likewise maximal for the most massive configuration \cite{Alford2012}. This generic behavior of the arising factor that incorporates the dependence on the particular star configuration is shown in Figure \ref{fig:configuration-dependence} for an APR EoS \cite{Akmal1998} and confirms that it is away from the mass limit indeed a monotonously decreasing function of mass. Therefore the uncertainty is in general much weaker than equation~(\ref{eq:saturation-amplitude-bound}) suggests and a strict limit is obtained for the minimum mass corresponding to the minimum $\tilde J$ and  the maximum radius for a given equation of state. The most strongly constrained and lowest masses have been observed in double neutron star binaries, which very likely present birth mass since no recycling is possible in such systems. These lie around $M\approx 1.3\,M_\odot$ \cite{Ozel2016}, and this value presents therefore a lower bound for the neutron star masses in millisecond pulsars which have a long accretion history and should therefore have masses significantly larger than their birth mass. 
Present observations constrain the radius of a $1.4\,M_\odot$ neutron star to $10\,{\rm km} \lesssim R_{1.4} \lesssim 11.5\,{\rm km}$ \cite{Ozel2016}. In this low mass regime the radius is very weakly mass-dependent anyway and using for a strict bound nevertheless $10\,{\rm km}$, as the lower limit of the above range, as well as the universal lower bound on $\tilde J$ equation~(\ref{eq:J-tilde-bound}) in equation~(\ref{eq:saturation-amplitude-bound}) gives therefore a rigorous amplitude bound. Taking further into account that the dominant relativistic corrections increase the r-mode frequency factor $\chi>1$ \cite{Idrisy:2014qca} gives finally
\begin{align}
\alpha_{{\rm sat}}&< 1.0\times10^{-9} \left(\frac{T_s}{10\,{\rm eV}}\right)^{2} \left(\frac{500\,{\rm Hz}}{f}\right)^{4} \nonumber \\
&\qquad\qquad\qquad \times \left[ \left(\frac{20 \pi \tilde J M}{1.3\,M_{\odot}}\right)^{-1}  \left(\frac{R}{10\,{\rm km}}\right)^{-2} \right]
\label{eq:numeric-saturation-amplitude-bound}
\end{align}
where the general bound is obtained when the bracket containing the dependence on the source properties is one, and a more restrictive bound can be obtained if information on source properties is available.

As discussed previously in \cite{Schwenzer}, in addition to the temperature dependence of these bounds, the other key parameter is the spin frequency, and fast spinning sources lead to significantly lower limits. If fast cooling processes are absent, equation~(\ref{eq:saturation-amplitude-bound}) is a good approximation for most millisecond pulsars since when expressed in terms of the core temperature the power-law exponents for photon- ($\theta_{\gamma}\sim 2$) (see eq. (\ref{eq:core-surface-relation}) below) and neutrino-emission ($\theta_{\nu} > 6$) are very different. In particular the contribution of  modified Urca processes ($\theta=8$) is suppressed by more than an order of magnitude already at about $5\times10^{5}\,{\rm K}$, i.e. at half the temperature where they are of equal size as photon emission. Yet, if fast cooling processes are present they can generally compete with photon cooling in observed sources.

\begin{figure}
	\includegraphics[scale=0.68]{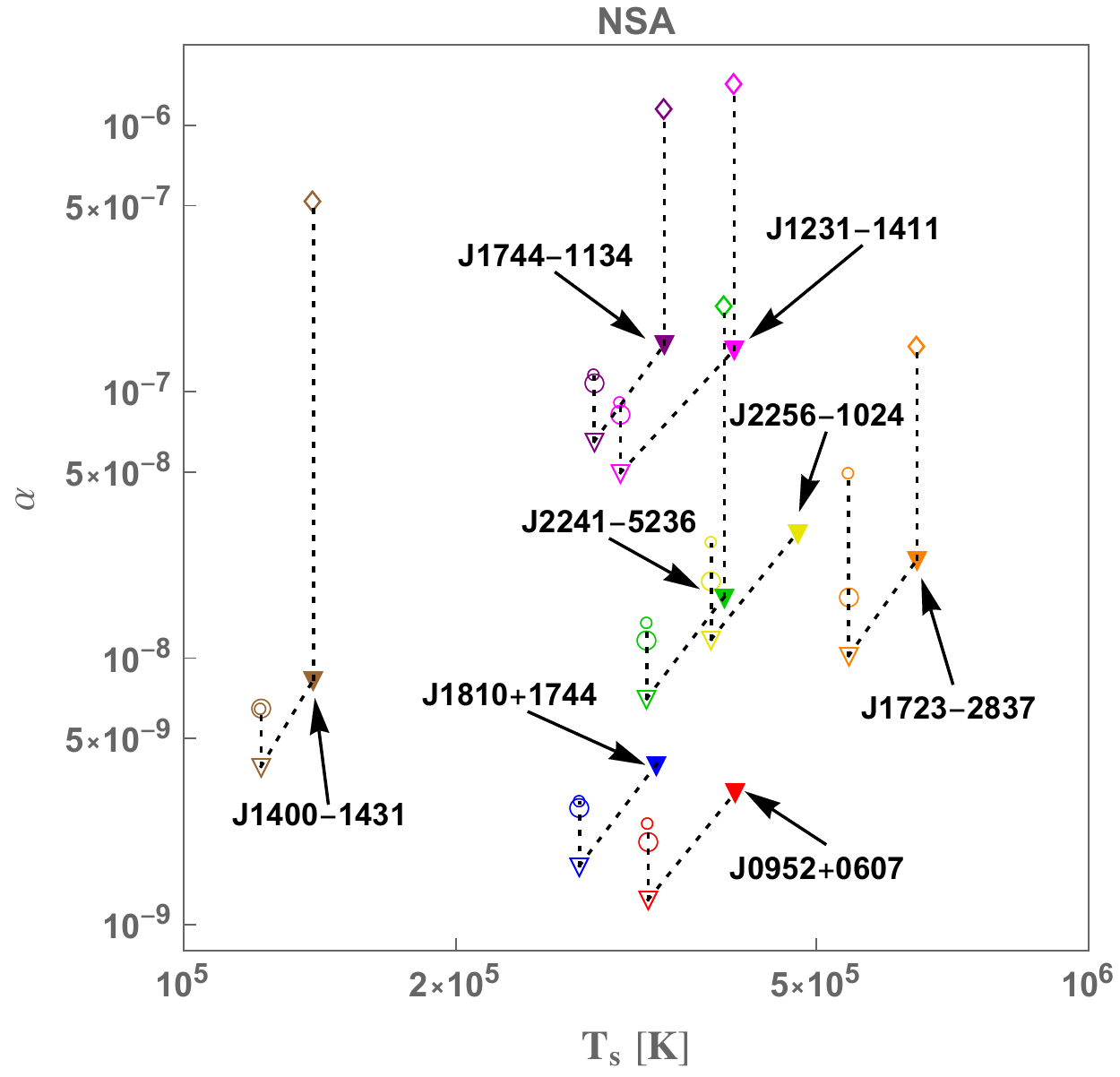}\caption{\label{fig:amplitude-bounds-NSA}Bounds on the r-mode amplitude stemming from thermal X-ray data (triangles) using a NSA model compared to those from pulsar timing data (diamonds) for the sources discussed in section \ref{sec:source}. Full triangles denote the rigorous upper limits and open triangles the bounds for fiducial standard assumptions on source properties. The circles show model computations for the special case that fast cooling would be present in the source, where the large and small ones correspond to the extreme cases of a fully catalyzed iron crust respectively a fully accreted light element envelope.}
\end{figure}

The rigorous bounds for the sources with new thermal X-ray data, employing the NSA model, are shown in Figure~\ref{fig:amplitude-bounds-NSA} (colored solid triangles) and the corresponding numerical values are given in Table~\ref{t_temp_lim}. Like for all the other uncertainties, in cases where different distance estimates have been given we always use the one that leads to the weakest bound.
Where timing data is available the data points are compared to their spindown limits (diamonds). As can be seen the spectral bounds are significantly below the spindown limits and those of some millisecond pulsars set very tight bounds on the r-mode amplitude\textemdash the most restrictive bound $\alpha_{{\rm sat}}\lesssim 3.1\times10^{-9}$ being obtained for the pulsar PSR J0952$+$0607, which reduces even to $\alpha_{{\rm sat}}\lesssim 1.2\times10^{-9}$ for fiducial source parameters, see below. However, also the bound from the new source PSR~\pulsarone~
for which we analyzed dedicated new XMM-Newton observations in this work, is not much weaker  and clearly below the results of our previous study \cite{Schwenzer}. Our lowest rigorous bound, obtained for PSR J0952$+$0607, is comparable to the value given for this source in \cite{Ho:2019myl}. In \cite{Bhattacharya2017} a similarly low bound of $\alpha \lesssim 2.5 \times 10^{-9}$ has been given based on a luminosity bound of 47 Tuc aa. However, this bound does not systematically take into account the uncertainties in the analysis.
In \cite{Chugunov:2017} the authors analyzed archival spectral results for selected sources as well as luminosity bounds, obtaining temperature bounds as low as $T_{\rm eff}^{\inf} \lesssim (3-4)\times 10^5\,{\rm K}$, as well as general expressions for the r-mode amplitude in line with \cite{Mahmoodifar2013,Schwenzer}, but refrained from giving bounds for individual sources.  
Similar bounds were also obtained for PSR J1640$+$2224 and PSR J1709$+$2313 in \cite{Mahmoodifar2017}, which are comparable to some of our less constraining sources due to the weaker surface temperature constraints of these sources. In general, the bounds from millisecond pulsars are by now also substantially lower than the best bound stemming from LMXBs \cite{Mahmoodifar2013}.

The assumptions entering the rigorous bounds are probably unrealistically conservative and if we use conventional source assumptions, as done in other studies, these bounds would be even tighter. 
A fiducial $1.4\,M_\odot$ APR neutron star \cite{Alford2012a} is e.g. given by the open triangles in Figure~\ref{fig:amplitude-bounds-NSA}~and in Table~\ref{t_temp_lim}, which yields values that can be more than a factor two lower. A better understanding of the equation of state of dense matter, e.g. due to advancements owing to the NICER observatory or future gravitational wave measurements, would therefore tighten these bounds. Moreover, as discussed, all these sources are expected to be substantially heavier than the mass limit of $M \gtrsim 1.3\,M_\odot$ assumed here. To reach their high frequencies, the millisecond sources we study need to have been spun up by accretion in a binary. The mass transfer should therefore result in star masses in the upper range of possible mass values, as is clearly observed for the few millisecond sources were mass measurements are available \cite{Ozel2016}. Unfortunately, to our knowledge none of the sources presented in Section \ref{sec:source} has currently a mass measurement \cite{Antoniadis2016}. Such independent mass constraints significantly enhance the bounds, as was e.g. shown for the case of PSR~J1023+0038 in \cite{Schwenzer}.

\begin{figure}
\includegraphics[scale=0.68]{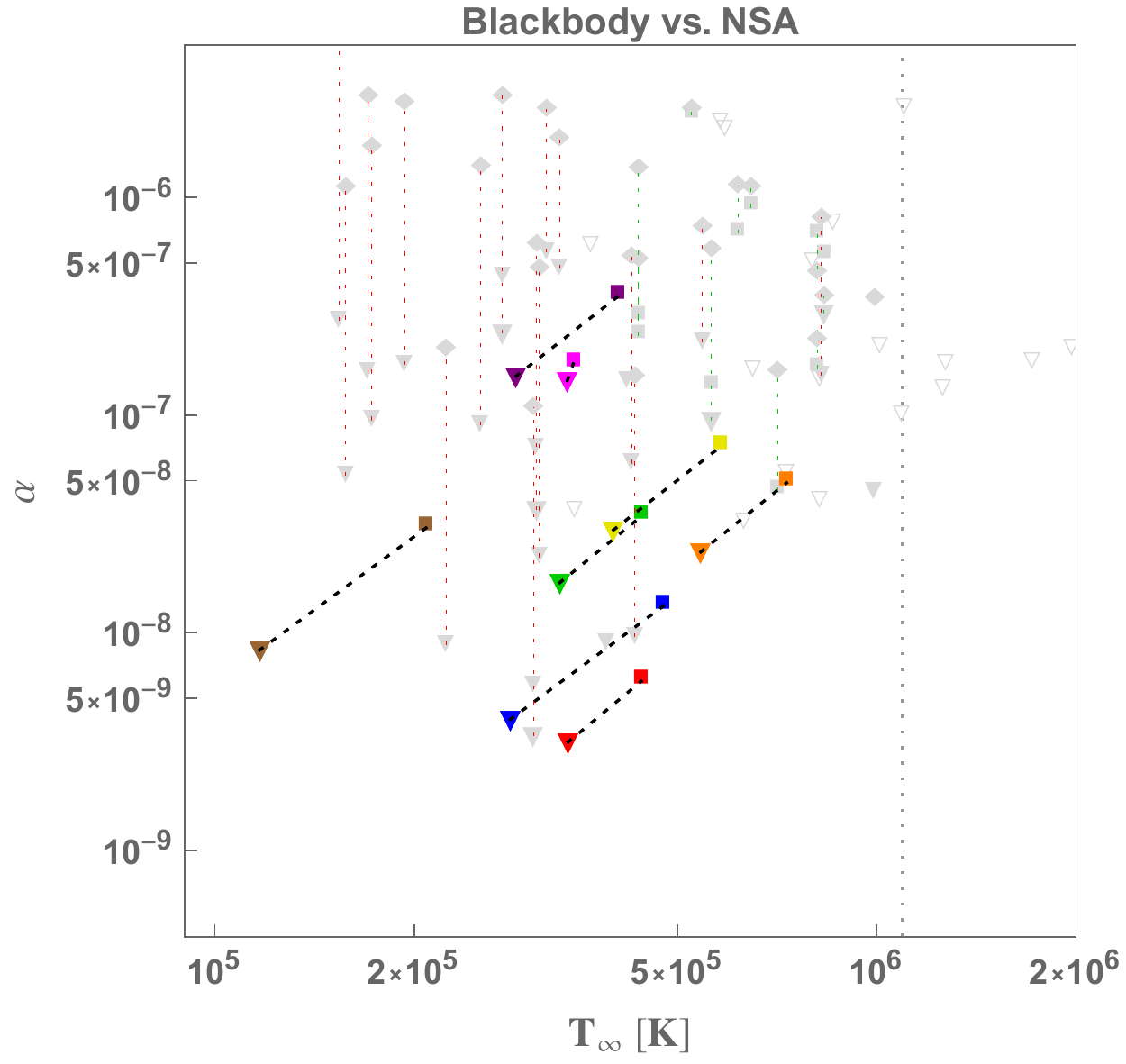}\caption{\label{fig:amplitude-bounds-BB}Comparison of the amplitude bounds when employing the NSA atmosphere model (large colored triangles, same colors as in Figure \ref{fig:amplitude-bounds-NSA}) vs. a simple {\em blackbody} model (small colored squares). The plot also shows our previous blackbody bounds (gray symbols) for the sources that had been studied (Schwenzer et al. (2017)).}
\end{figure}

In Figure~\ref{fig:amplitude-bounds-BB} our bounds using the more realistic NSA model (triangles) are compared to the previously employed {\em blackbody} model (squares), which shows that the new bounds based on the NSA model are significantly stronger, owing to the tighter thermal bounds. As noted before in \cite{Chugunov:2017}, this is a generic effect that is clearly seen for most of the sources. However, when the spectral shape of the data for a particular source differs strongly from that of one of the two models this can result in a poorer fit and a weaker deviation. For comparison Figure~\ref{fig:amplitude-bounds-BB} also shows {\em blackbody} results for other sources taken form our previous analysis \cite{Schwenzer}. Most of these are mere luminosity bounds that did not rely on detailed spectral fitting of a surface component. As can be seen our new results significantly surpass these previous bounds from values around $10^{-8}$ to bounds not far above $10^{-9}$. However, taking into account that the NSA model generically strengthens the bounds, there are several sources, for which a proper spectral reanalysis with the NSA model or even the use of additional improved data would be very promising. 

As discussed, the present bounds were obtained for the standard case that photon emission from the atmosphere dominates the cooling of the star. This is the case when merely standard neutrino cooling mechanisms, like modified Urca emission, are present. However, fast cooling mechanisms,
like pair breaking emission or even direct Urca processes, could be present in these sources. In this case, even at the observed low temperatures, neutrino emission from the bulk would still dominate. As noted in \cite{Mahmoodifar2013,Schwenzer} this could in principle lead to slightly weaker bounds. These depend on the particular neutrino cooling mechanism involving additional microscopic physics and it is therefore harder to give a rigorous expression. However, as done previously \cite{Mahmoodifar2013,Schwenzer} one can get an idea of the maximum impact of fast cooling by considering the extreme case of hadronic direct Urca cooling, presenting the fastest know cooling mechanism. In this case the bound reads \cite{Schwenzer}
\begin{equation}
\alpha_{{\rm sat}}\leq\sqrt{\frac{3^{8}5^{2}\tilde{L}\Lambda_{{\rm QCD}}}{2^{15}\tilde{J}^{2}\Lambda_{{\rm EW}}^{4}GM^{2}R^{3}}}\frac{T^{4}}{\Omega^{4}} \label{eq:direct-Urca-bound}
\end{equation}
where the dimensionless quantity $\tilde{L}$ characterizes the neutrino emission in the star and $\Lambda_{{\rm QCD}}\!=\!1\,{\rm GeV}$ and $\Lambda_{{\rm EW}}\!=\!100\,{\rm GeV}$ are generic normalization scales.
We use a maximum mass neutron star model with an APR equation of state ($M\approx 2.2\, M_\odot$, $\tilde L\approx 2.3\times 10^{-5}$), that features a sizable core where direct Urca emission is allowed \cite{Alford2012a}. Such a massive and compact source would according to equation~(\ref{eq:direct-Urca-bound}) impose a stronger bound and therefore we correct for the deviation from our fiducial parameter set from a $1.4\, M_\odot$ star discussed above. This way we take into account that for another equation of state direct Urca processes might already be possible in a $1.4\,M_\odot$ source. Since eq.~(\ref{eq:direct-Urca-bound}) depends on the core temperature, a crust model is required and we consider the extreme cases of a fully catalyzed iron crust \cite{Gudmundsson1983} and a fully accreted light element envelope \cite{Potekhin1997} to gauge its impact.
The results are given by the large (catalyzed) and small circles (accreted) in Figure~\ref{fig:amplitude-bounds-NSA} and as can be seen even with these extreme assumptions the obtained bounds are roughly comparable to those stemming from pure photon emission. Only for the hottest sources, where the bounds are poor anyway, an accreted light element crust does significantly weaken the bounds if fast cooling processes are present.

The obtained amplitudes are not too far from the regime $\lesssim10^{-11}-10^{-10}$ where r-modes could be completely ruled out in theses sources, since even the fastest millisecond pulsars could cool out of the instability region \cite{Alford2013}. What is more is that these amplitudes are many orders of magnitude below those that well established saturation mechanisms, like for instance mode-coupling \cite{Bondarescu2013}, can provide. 
Therefore, neutron stars with minimal damping, i.e. a homogenous, non-superfluid star made of standard npe$\mu$-matter\footnote{Note that our definition of a "minimal" neutron star composition is more conservative than in the minimal cooling scenario, in that it does not include superfluidity.}, require another strong non-linear dissipation mechanism that can completely damp or saturate r-modes at such low amplitudes. Such an additional mechanism is not established at this point in ordinary neutron matter and therefore the minimal picture of neutron stars is currently incompatible with the astrophysical data. This points to fascinating new physics and there are several interesting proposals \cite{Madsen1999,Alford2013,Alford2014}.

The bounds on the r-mode amplitude impose finally corresponding bounds on the gravitational wave signal from these sources. As shown in \cite{Alford2014,Schwenzer}, despite the large spin frequencies of these sources, even the weaker previous amplitude bounds led to a gravitational wave strain for these sources that would be too low to be detectable with current gravitational wave detectors. The strengthened bounds underline this conclusion and show that a potential r-mode emission of known millisecond pulsars is unfortunately out of reach at present. Young sources present therefore a far more promising target to detect gravitational wave emission due to r-modes \cite{Alford:2014}.

\section{Presence of r-modes}\label{sec:prermode}

It had been shown, that, if millisecond sources are ordinary neutron stars with known damping mechanisms, they would be trapped within the r-mode instability region \cite{Alford2013}, This holds since we on one side observe their tiny spindown rates and on the other side the r-mode amplitudes obtained from well constrained saturation mechanisms would strongly heat these sources so that they cannot cool out of the instability region. However, the increasingly restrictive temperature bounds, imposing corresponding bounds on the r-mode amplitude in millisecond sources, also open the possibility that r-modes can be completely absent in many sources even in case a minimal neutron star scenario is realized. As shown in \cite{Alford2013} for amplitudes as low as $\alpha \lesssim 10^{-11} - 10^{-10}$ the r-mode heating would be small enough that all of these very old sources should have by now cooled out of the instability region. R-modes and the accompanying gravitational wave emission are therefore absent and no saturation mechanism is required at present. Yet, as argued in \cite{Alford2013} this would nevertheless have required a humongous dissipation in their interior during their evolution since it is known from observed LMXBs that they start their cooling evolution after recycling right within the instability region of a minimal neutron star \cite{Haskell:2012,Haskell2015IJMPE}. 

In this section we discuss the possibility that some sources are actually outside of the instability region in detail and take into account the uncertainties in the analysis to derive a simple but rigorous condition based on X-ray observations for when r-modes will be absent in a millisecond pulsar. This information should be useful for future continuous gravitational wave searches. Despite being out of reach of current detectors, and except for more complicated evolutionary scenarios, see e.g. \cite{Kantor:2015xsj}, millisecond pulsars undergoing r-mode oscillations could be ideal sources for gravitational wave astronomy due to their enormous stability. They would allow us to perform precision multi-messenger gravitational wave observations over long time intervals that could determine various source properties ranging from bulk observables, like mass, radius and moment of inertia, to thermal properties and even independent distance measurements that are not affected by interstellar absorption \cite{Aasi2013,Alford2014,Kokkotas2016}.

General analytic estimates for the boundary segments of the r-mode instability region had been obtained in \cite{Alford2010}. In the low temperature region, relevant for millisecond pulsars, shear viscosity is the relevant mechanism, which in neutron matter (as well as in most other forms of matter) dominantly stems from the scattering of relativistic particles due to long-ranged interactions mediated by Landau-damped transverse gauge bosons \cite{Shternin:2008es}. Both in neutron matter and non-CFL quark matter electromagnetic electron scattering is the relevant mechanism, whereas in the quark case there are analogous additional contributions from quark scattering, reducing the mean free path and increasing the instability region. Generalizing the result given in \cite{Alford2010} by taking into account a general r-modes dispersion relation $\omega = - 4/3 \chi \Omega$ via the correction factor $\chi \approx 1$, the low-temperature segment of the instability boundary is in terms of the core temperature $T$
\begin{equation}
\Omega_{\rm ib}=1.12\frac{\tilde{S}^{1/6}\Lambda_{{\rm QCD}}^{7/9}}{\left(3-2\chi\right)^{1/6}\chi^{5/6}\tilde{J}^{1/3}G^{1/6}M^{1/3}R^{1/2}} T^{-5/18} \label{eq:y}
\end{equation}
where $\tilde J$ and $\tilde S$ are dimensionless constants describing weighted averages of the energy density and of the shear viscosity over the entire source. This requires to connect the core temperature to the observed surface temperature. Taking into account that the thermal conductivity in the core of a neutron star is very large, it has been shown \cite{Gudmundsson1983} that the core temperature $T$ is to a very good approximation a function of the single quantity $T_s^4/g_s$, where $g_s\!\equiv\!GM/(R\sqrt{1-2GM/R})$ is the surface gravitational acceleration.  This function is determined by the detailed heat transport in the crust 
and the connection stemming from numerical simulations can generally up to corrections at the 10\%-level be approximated by a simple Power-law. Since the millisecond sources considered here are currently not perceivably accreting and most of them likely have not done so for a very long time, easily exceeding hundreds of millions of years, a model of a catalyzed iron crust without light element admixtures is used, as will be discussed in more detail below. The Power-law dependence takes in this case the form \cite{Gudmundsson1983}
\begin{equation}
\frac{T}{10^8\,{\rm K}} \approx 1.288 \left(\frac{10^{14} \,{\rm g\, cm^{-2}}}{g_{s}} \left( \frac{T_{s}}{10^6\,{\rm K}}\right)^4 \right)^{0.455}
\label{eq:core-surface-relation}
\end{equation}
and a quantitatively very similar Power-law (prefactor 1.429, exponent 0.444) is obtained for the leading component in the independent analysis \cite{Potekhin1997} (where the accuracy of the fit was improved to the \%-level by adding a second, subleading Power-law component that contributes less than 3\% for $T\gtrsim 10^8\,{\rm K}$). Quantitatively comparing the two results shows that they deviate by less than 10\% for $T_s \gtrsim 10^5\,{\rm K}$ so that the core-surface temperature relation should be accurate at this level. 
Inserting this above the non Power-law correction factor $\sqrt{1-R_s/R}$, where $R_s=2GM$, enters via a very low Power $\approx 0.06$ and therefore amounts for the realistic regime of neutron star radii $R>2 R_s$ to less than 3 percent. Similarly in the relevant regime $1-\chi \ll 1$ the dependence on the r-mode dispersion relation can be simplified. 
and the result for the boundary of the r-mode instability boundary takes the form
\begin{align}
f_{\rm ib}=&\left(403\pm 40\right)\,{\rm Hz}\frac{\tilde{S}_{1.4}^{1/6}}{\chi^{1/2}\tilde{J}_{1.4}^{1/3}}  \nonumber \\
&\times \left(\frac{M}{1.4\,M_{\odot}}\right)^{-0.207}\left(\frac{R}{10\,{\rm km}}\right)^{-0.753}\left(\frac{T_{s}}{10^{5}\,{\rm K}}\right)^{-0.506} \label{eq:instability-frequency}
\end{align}
where $\tilde{J}_{1.4}$ and $\tilde{S}_{1.4}$ are the corresponding values for a standard $1.4\, M_\odot$ neutron star \cite{Alford2012a} with an APR equation of state \cite{Akmal1998}. 
In order to determine if r-modes could be completely absent in observed sources, we are interested in the maximal size of the instability region, that standard neutron stars without enhanced damping could have, i.e. the minimal possible frequency value equation~(\ref{eq:instability-frequency}) can take. As seen in Figure~\ref{fig:configuration-dependence} the product of factors encoding the source dependence is basically independent of the mass, with a shallow minimum around $2 M_\odot$. To get a bound we use here correspondingly a $2 M_\odot$ APR \cite{Akmal1998} model \cite{Alford2012a} and estimate the residual uncertainty due to the equation of state as $10\%$. The shear viscosity of dense neutron matter stems from sufficiently constrained leptonic processes \cite{Shternin:2008es} and the dimensionless factor $\tilde S$ should therefore be uncertain by at most a factor two. The uncertainty due to the r-mode dispersion is around $10\%$ and the small deviation of the temperature exponent from 1/2 imposes over the relevant temperature range merely a $1\%$ effect. 
This yields for a given temperature a minimal frequency to which the instability region can extend. A source spinning with a smaller frequency $f$ will even within the sizable uncertainties be outside of the instability region, so that it cannot emit gravitational waves due to r-modes. We obtain correspondingly finally the simple bound, that a source is stable if
\begin{equation}
f \leq \left(389^{+165}_{-116} \right) {\rm Hz} \; \sqrt{\frac{T_{s}}{10^{5}\,{\rm K}}} \label{eq:frequency-bound}
\end{equation}
As discussed, this expression determines if a source is even outside of the maximally possible instability region, obtained for a minimal neutron star composition. Any additional damping decreases the instability region, so that a given source will be even more stable to r-modes. For instance it is quite well established that neutron superfluidity and proton superconductivity are present in neutron stars. In this case the shear viscosity is enhanced \cite{Shternin:2008es,Shternin:2018}, yet the detailed impact on the instability region depends on superfluid parameters, like the critical temperature, that are still rather uncertain. Proton superconductivity would somewhat increase the (irrelevant) upper boundary of the uncertainty band in equation~(\ref{eq:frequency-bound}) (given here for a non-superfluid composition), but nevertheless could not stabilize the r-mode instability in the faster sources studied in this work.

Similarly, it might be that a source had either a rather recent accretion history. Moreover, it is likely that the crust cannot be completely catalyzed even long after accretion stopped \cite{Yakovlev:2006fi,Beznogov:2016ejn}. As shown in \cite{Potekhin1997} for a given surface temperature the internal temperature would be systematically lower for an accreted crust, compared to the catalyzed iron crust \cite{Gudmundsson1983} we use here. According to equation~(\ref{eq:y}), as a function of the surface temperature, the instability region would correspondingly lie at higher frequencies and therefore again be smaller. A source that fulfills equation~(\ref {eq:frequency-bound}) would therefore be even more stable to r-modes if its crust would contain light elements. All these considerations show that equation~(\ref {eq:frequency-bound}) indeed presents a rigorous condition for the absence of r-modes in observed sources. 

\begin{figure}
\includegraphics[scale=0.68]{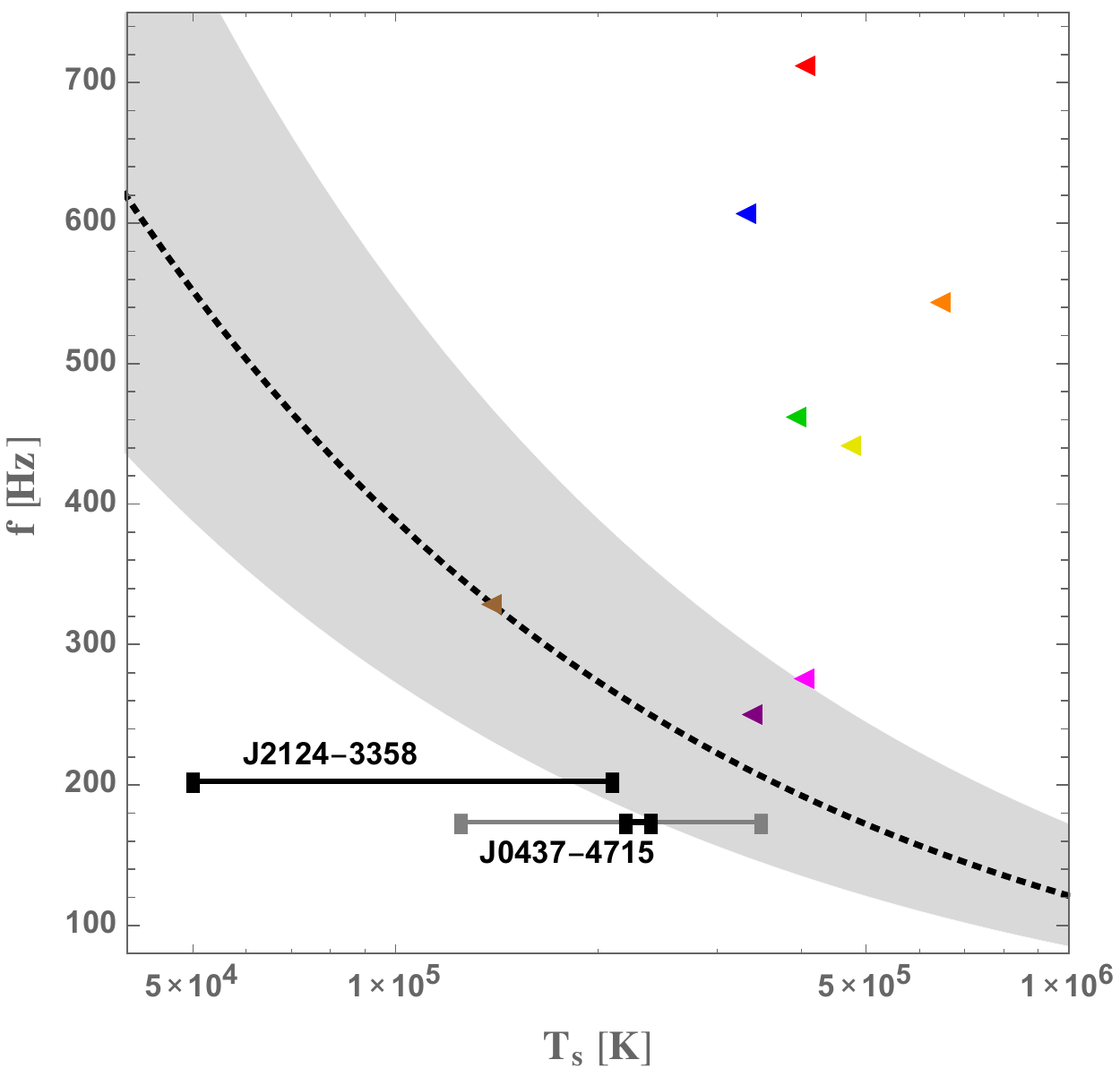}\caption{\label{fig:instability-region}Sources with spectral upper temperature bounds discussed in this work (triangles) compared to the analytic estimate for the boundary of the r-mode instability region equation~(\ref{eq:frequency-bound}) with uncertainty band. The color coding is the same as in Figure~\ref{fig:amplitude-bounds-NSA}. The horizontal lines show the uncertainty intervals for two sources with actual temperature measurements.}
\end{figure}

The comparison of the sources discussed in this work with the boundary of the instability region is shown in Figure~\ref{fig:instability-region}. As can be seen, within the sizable uncertainties none of the considered sources is clearly outside of the instability region of a neutron star with standard viscous damping mechanisms, yet. Therefore r-mode gravitational wave emission cannot be excluded at present. However some of them could be excluded in the future if tighter temperature limits can be obtained or the theoretical uncertainties on the instability region can be reduced. The figure also shows the two sources PSR J0437$-$4715 \cite{Durant:2011je,Gonzalez:2019} and PSR J2124$-$3358 \cite{Rangelov:2016syg} for which an actual surface temperature estimate could be obtained using additional UV observations. In case of PSR J0437$-$4715 the large gray line shows the result of a previous analysis \cite{Durant:2011je} and the narrow black segment the strongly improved error margins of a recent study \cite{Gonzalez:2019}.
As seen, even though within the sizable uncertainties the presence of r-modes cannot be completely excluded, they are likely outside of the instability region. 

At very low temperatures the shear viscosity of dense hadronic matter can be dominated by screened, short-ranged interactions \cite{Shternin:2008es} which reduces the instability region and correspondingly makes it even easier that fast spinning sources can be outside. This effect will have to be included once even lower temperature bounds or measurements are obtained.

\section{Discussion and Conclusion}\label{sec:discus}

Using new dedicated {\em XMM-Newton} observations of PSR \pulsarone~and PSR \pulsarnine~and refined X-ray spectral analyses of several other millisecond pulsars, we set new bounds on the surface emission of these sources. We obtained our lowest limit for PSR~\pulsarsix, which equals the lower limit of the temperature limit where the NSA model is valid, as defined in Xspec. This is a remarkable limit, as it hints at a surface temperature lower than 10$^{5}$~K. At such lower temperature values the fully ionized Hydrogen atmosphere assumption may not be valid anymore. Apart from PSR~\pulsarsix, our upper bounds are in general found to be around 25~eV. The exact value for a neutron star, obviously depends on the continuum emission, the distance, hydrogen column density in the line of sight and the instrument being used. Regarding the latter point, it is obvious that instruments sensitive to far ultraviolet are necessary. The proposed LUVOIR mission may eventually contribute a lot to constraining the surface emission of these old neutron stars \citep{2018luvoir}. The problem at such low wavelengths though is that at these regions the companion object may be so dominant in the total emission of the system that it may be impossible to constrain the surface temperature of the neutron star (see however, \cite{Rangelov:2016syg}). Therefore one other option would be to discover more nearby millisecond pulsars, and observe them in the soft X-ray band. The recently launched eROSITA may provide critical contributions \citep{Merloni2012}.

Since r-modes would strongly heat such sources, our results impose tighter bounds on the r$-$mode amplitude than obtained in previous analyses, as well as obtained from accreting sources in LMXBs or mere spin down limits. The key finding of r-mode astero-seismology has been that at temperatures present in LMXBs, standard dissipative mechanisms in non-superfluid neutron stars cannot damp r-modes in sources spinning faster than $f\gtrsim 100\,{\rm Hz}$ \cite{Lindblom1998}, \cite{Alford:2014}~(see Table \ref{t_msp_lit}) ---and slightly higher values are obtained for minimal superfluid compositions, where proton superconductivity can increase the damping \cite{Shternin:2008es,Shternin:2018}. Under the additional assumption of spin equilibrium, this conclusion is also directly obtained from the accretion luminosity in LMXBs \cite{Ho:2011PhRv}.
For a faster-spinning star made of neutron matter, this therefore requires a mechanism that eventually saturates the mode due to a nonlinear, amplitude-dependent enhancement of the dissipation, or dynamically in the course of their evolution \cite{Andersson2001,Reisenegger2003,Chugunov:2017}. 
Several such mechanisms have been proposed, but our new rigorous bounds  $\alpha\lesssim 3\times 10^{-9}$ are by now so far below the saturation amplitudes $\alpha\sim10^{-6}-10^{-5}$ that standard mechanisms in neutron stars \cite{Bondarescu2013} might provide, that it becomes more and more unlikely that the saturated r-mode scenario is realized \cite{Haskell2015}. An exception is a hybrid star with a sharp interface, where phase conversion dissipation, presenting the strongest known dissipation mechanism \cite{Alford:2014jha}, can provide even lower saturation amplitudes. 

Correspondingly our results show that there must be significant additional damping in these sources and further enlarge the discrepancy between standard, well-constrained damping mechanisms and the astrophysical data. 
Therefore, a minimal neutron star composition is by now basically ruled out by the astrophysical data, and it is likely that there is actually a mechanism, that does not merely saturate r-modes at very low amplitudes, but that completely damps them away. This requires very strong dissipation that could either stem from exotic phases of matter with inherently strong dissipation \cite{Alford:2019oge} or from the structural complexity of the star. 


The first case is simpler since it is described by the local hydrodynamic dissipation coefficients, which are generally determined by the dynamics of the low energy degrees of freedom of the corresponding phase. An important example is ungapped quark matter, where the bulk viscosity due to non-leptonic, flavor-changing interactions is resonantly enhanced under the conditions in cold neutron stars, since the time scale of the weak processes matches the time scale of r-mode oscillations \cite{Alford:2014}. Color-superconductivity in general significantly suppresses the dissipation \cite{Alford:2019oge}. Goldstone bosons could become relevant if their mean free path becomes large. However, at compact star temperatures in CFL quark matter it even largely exceeds the size of the star so that Goldstone bosons behave ballistically \cite{Andersson:2010sh}, which complicates the analysis preventing so far precise quantitative predictions.

The simplest example for a dissipative mechanism due to the structural complexity of the star is the Ekman-layer rubbing of the fluid in the core along a solid crust \cite{Lindblom:2000gu}. Yet, even under most favorable assumptions this cannot provide the required damping \cite{Alford2014}.
Further examples for dissipation in structurally complex sources are given by a hadronic-quark interface \cite{Alford:2014jha}, non-uniform ("pasta") phases \cite{Horowitz:2008vf}, extended field configurations like fluxtubes and/or vortices \cite{Haskell:2013hja}, or the interactions of the r-mode with more localized oscillation modes, e.g. in the superfluid core \cite{Gusakov:2013jwa,Ho:2019myl}. All these are more complicated to describe since they require detailed model assumptions about the poorly constrained structural composition of neutron stars and therefore only allow very rough estimates for the possible dissipation. 

As had been shown in \cite{Alford2013}, at very low saturation amplitude, slower spinning MSPs\textemdash that would be trapped inside the instability region if the r-mode amplitude would be larger\textemdash can actually cool out of the r-mode instability region on times much shorter than their billion year ages, and therefore r-modes are not expected in these sources. We derived an analytic expression for the limiting spin-frequency, below which r-modes can be ruled out in a given source. This also allows us to estimate the involved uncertainties and we find that within the sizable uncertainties r-modes cannot be ruled out at present in the considered sources. 

\section*{Data availability}

The datasets were derived from sources in the public domain: XMM-Newton (http://nxsa.esac.esa.int/nxsa-web/\#search) and Chandra (https://cda.harvard.edu/chaser/).

\section*{Acknowledgements}

We thank the anonymous referee very much for the detailed suggestions, which clearly improved the manuscript. We thank Paul Ray who brought PSR~\pulsarsix~ to our attention and Sebastien Guillot for very helpful comments. This research is supported by TUBITAK via the project number 117F312. T.G. is supported in part by the Royal Society-Newton Advanced Fellowship program, NAF\textbackslash R2\textbackslash180592. T.G. and T.B. are supported in part by the Turkish Republic Ministry of Development with the project number, 2016K121370.
Partly based on observations obtained with {\em XMM-Newton}, an ESA science mission with instruments and contributions directly funded by ESA Member States and NASA. 
The scientific results reported in this article are based in part on observations made by the CXO.
This research has made use of data and/or software provided by the High Energy Astrophysics Science Archive Research Center (HEASARC), which is a service of the Astrophysics Science Division at NASA/GSFC and the High Energy Astrophysics Division of the Smithsonian Astrophysical Observatory.





\end{document}